\documentclass{article}
\usepackage[superscript]{cite}
\usepackage[english]{babel}
\usepackage{graphicx}
\usepackage{subcaption}
\usepackage{float}
\usepackage[export]{adjustbox}
\usepackage{enumitem}
\usepackage{multirow}

\usepackage[letterpaper,top=2cm,bottom=2cm,left=3cm,right=3cm,marginparwidth=1.75cm]{geometry}

\usepackage{amsmath}
\usepackage{amssymb}
\usepackage{amsfonts}
\usepackage{authblk}
\usepackage[superscript]{cite}

\usepackage[colorlinks=true, allcolors=blue]{hyperref}

\title{Orientation and mobility test in virtual reality, a tool for quantitative assessment of functional vision: dataset and evaluation in healthy subjects}

\author[1]{Yujie Huang\textsuperscript{*,\dag}}
\author[1,2]{Audrey Crozet\textsuperscript{*}}
\author[1]{Toinon Vigier}
\author[1]{Alexandre Bruckert}
\author[1,3]{Patrick Le Callet}
\author[1,2]{Pierre Lebranchu}

\affil[1]{Nantes Université, École Centrale Nantes, CNRS, LS2N, UMR 6004, F-44000 Nantes, France}
\affil[2]{CHU Nantes}
\affil[3]{Institut universitaire de France (IUF)}

\begin{document}
\maketitle
\let\thefootnote\relax
\footnotetext{\textsuperscript{*}These authors contributed equally to this work.}
\footnotetext{\textsuperscript{\dag}Corresponding author: Yujie Huang (\texttt{yujie.huang@etu.univ-nantes.fr})}

\begin{abstract}
\textbf{Purpose:} The purpose of this study was to develop and evaluate a novel virtual reality seated orientation and mobility (VR-S-O\&M) test protocol designed to assess functional vision. This study aims to provide a dataset of healthy subjects using this protocol and preliminary analyses.

\textbf{Methods:} We introduced a VR-based O\&M test protocol featuring a novel seated displacement method, diverse lighting conditions, and varying course configurations within a virtual environment. Normally sighted participants (N=42) completed the test, which required them to navigate a path and destroy identified obstacles. We assessed basic performance metrics, including time duration, number of missed objects, and time before the first step, under different environmental conditions to verify ecological validity. Additionally, we analyzed participants' behaviors regarding missed objects, demonstrating the potential of integrating behavioral and interactive data for a more precise functional vision assessment.

\textbf{Results:} Our analyses revealed that performance metrics, including time duration and time before the first step, were affected by lighting levels, with a threshold effect observed at higher levels. Although no significant differences were found in time duration and orientation time across different course configurations, the number of missed objects varied significantly, indicating differences in difficulty levels. In addition, behavioral analyses identified different several potential error categories (e.g., lack of visual analysis of the environment, difficulty in detection, misuse of the setup) and the potential most challenging object features (position and color) to infer vision problems. 

\textbf{Conclusions:} Our VR-S-O\&M test protocol demonstrated feasibility and validity on healthy subjects, showing the impact of illumination, revealing the exist of different difficulty levels in our environment configurations. This study highlighted the need for further investigation into the causes of errors and the effects of different configurations on performance.  

\textbf{Translational Relevance:} Our VR-S-O\&M test protocol, along with the first O\&M behavior dataset, presents significant opportunities for developing more refined performance metrics for assessing functional vision and enhancing the quality of life.

\end{abstract}

\section{Introduction}
Functional vision is a key concept in visual impairment and refers to a person's ability to perform vision-related tasks\cite{dutton2001cerebral}. A decline in functional vision can significantly affect the Quality of Life (QoL) across all age groups\cite{schmetterer2023endpoints}.
In contrast to visual functions such as visual acuity, color vision, contrast vision, ocular motility and visual field which are assessed using standardized, validated and quantitative instruments, the evaluation of functional vision is far more complex and less commonly performed. 

Currently, there are two primary methods for measuring functional vision\cite{bennett2019assessment, silveira2019exploring}: quality of life questionnaires\cite{finger2014developing, goldstein2022nei, jeanjean2020validation, leske2021quality, csahli2021comparison, wolffsohn2000design} and physical orientation and mobility (O\&M) tests\cite{chung2018novel, geruschat2012orientation, hassan2002vision, kumaran2020validation, nau2014standardized}. The former lacks objectivity, while the latter presents challenges in distribution, reproducibility, and is often bulky, time-consuming and costly. One of the most advanced O\&M tests for assessing functional vision is the Multi Luminance Mobility Test (MLMT)\cite{chung2018novel}. This test was specifically developed to demonstrate the functional benefit of treatment voretigene neparvovec, the first reimbursed gene therapy in ophthalmology, for patients with hereditary retinal dystrophy caused by the bi-allelic mutation of RPE65. The MLMT assesses functional vision by incorporating standardized variations in light conditions.
Reproducibility was a key consideration in the development of MLMT, ensuring that different test centres could conduct the tests under identical conditions. 
To minimize the learning effect, multiple courses with obstacles were designed. Each run was recorded and reviewed by two independent, trained observers. The test demonstrated clinical benefits in functional vision by showing that patients had shorter course completion times and fewer collisions with obstacles after treatment. However, despite its thoughtful design, the MLMT has several limitations, including the need for a dedicated room with calibrated lighting, the time-consuming nature (need for the assessor to change the layout of the course frequently), the cost and the constraints imposed by its patent. Many of these drawbacks can be overcome by using virtual reality (VR). Virtual reality is a rapidly advancing field that is already being used in ophthalmology for low vision rehabilitation\cite{bowman2017individuals, iskander2021virtual, pur2023use}. Building on the MLMT, virtual reality orientation and mobility tests have been developed to assess functional vision in patients with inherited retinal dystrophies (IRD) \cite{aleman2021virtual, bennett2023optimization}, to simulate the vision with a retinal prosthesis and optimize its placement on the retina\cite{zapf2015assistive, zapf2016assistive}, as well as to evaluate the severity of visual impairment in glaucomatous patients\cite{lam2020use}. The primary limitation of these virtual reality O\&M tests is that they require the subject to move physically, which necessitates a large space, poses potential safety risks, induces fear, and is not suitable for patients with mobility impairments.

We believe that virtual reality can address the primary limitations of the aforementioned tests, particularly by allowing patients to remain seated in a swivel chair during the assessment. Visually impaired patients often have difficulty walking, which may be related to a fear of falling or bumping into obstacles\cite{white2015fear}. By opting for a seated position, we can reduce the variability in walking patterns across different patients and enable those with motor impairments (such as multiple sclerosis, diabetes, or stroke) to undergo mobility testing. 
With the development of innovative therapies and medical devices for visually impaired individuals, there is an increasing need for a rigorous, quantitative, and reproducible assessment of functional vision. The key advantages of virtual reality include the ability to create controlled virtual environments that closely mimic real-world conditions, the reproducibility of testing environments, the automated recording of attentional and behavioral data, and the potential reduction in costs. Assessing visual functions alone is not always sufficient to demonstrate clinical therapeutic benefits in ophthalmology. This is partly because individual visual functions do not fully capture the complexity of functional vision in everyday life, and partly because existing tools for assessing different visual functions may not be feasible for patients with visual impairments. In addition to its utility in clinical research, this tool could also be valuable for studying the natural history of blinding conditions.

In this study, we introduce our Virtual Reality Seated Orientation and Mobility dataset (VR-S-O\&M) involving healthy subjects, and provide analyses of its validation. Specifically, we describe the development of our O\&M test, the details of the displacement mode, the various course configurations and lighting levels, and the evaluation conducted with visually healthy subjects. To our knowledge, this is the first dataset for O\&M testing of functional vision that includes rich behavioral information and detailed scenario configurations, offering researchers a promising opportunity to better understand individual behaviors.

\section{Methods}
\subsection{Materials}
The virtual reality environments were developed using Unity, coded in C\# (Unity editor version 2020.3.35f1, SteamVR version 1.24.6, SRanipal Runtime version 1.3.6.8). The Head-Mounted Display (HMD) used was the HTC Vive Pro Eye HTC (Corp., Xindian, New Taipei, Taiwan). This device was specially chosen for its built-in Tobii eye tracking system. The headset features a display resolution of $1440 \times 1600$ pixels per eye, 90 Hz refresh rate and a Field of View (FoV) of 110$^{\circ}$ as stated by the manufacturer. Recent work\cite{sauer2022assessment} found the official FoV specifications for most VR headsets to be overstated and reports the effective FoV for the Vive Pro Eye at 94$^{\circ}$ horizontally. The eye tracking system equipped in this HMD runs at 120 Hz sampling rate and reports to achieve a spatial accuracy of 0.5$^{\circ}$ - 1.1$^{\circ}$ (HTC Corporation, 2021).

\subsection{Test Design}
\begin{figure}[H]
    \centering
    \includegraphics{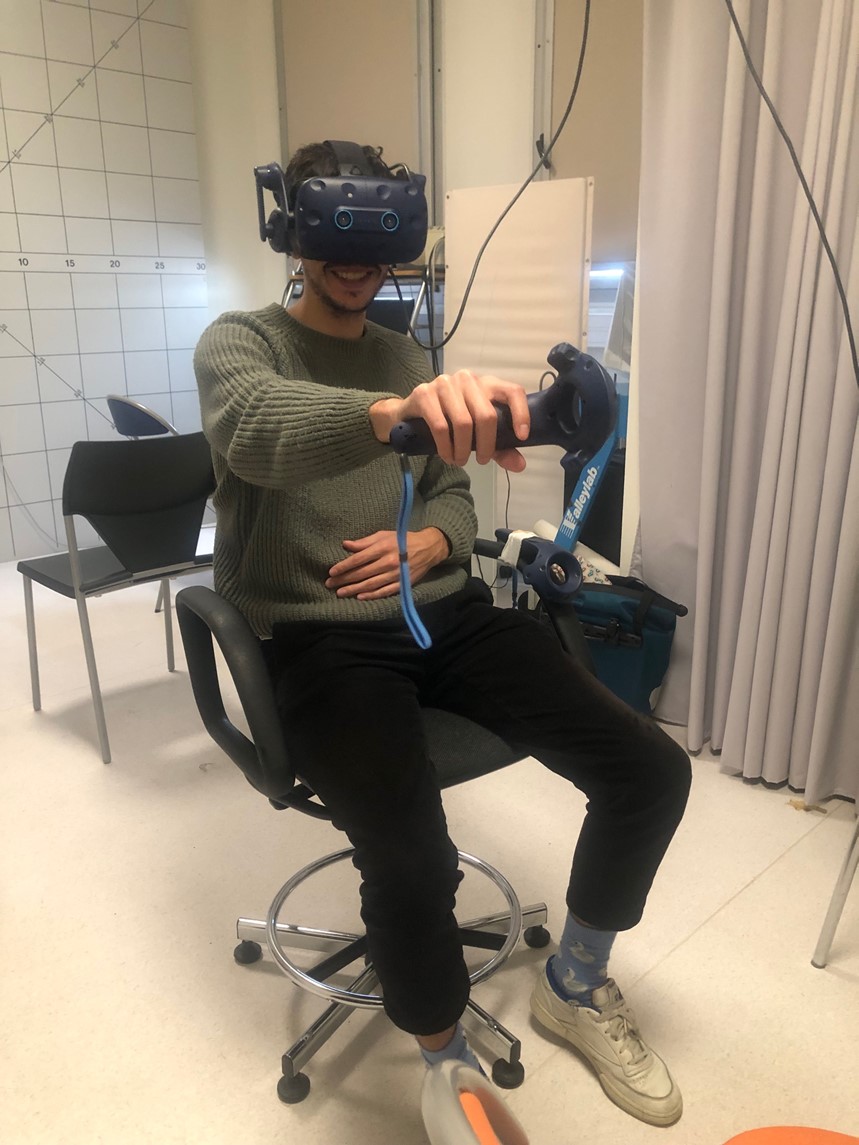}
    \caption{A subject doing the VR-S-O\&M test with his written consent.}
    \label{fig:subject}
\end{figure}

The goal for participants in our VR-S-O\&M test is to follow a path on the floor and destroy objects along the way to ensure they have been detected in the virtual environment. Figure \ref{fig:subject} shows a participant performing the test. During the test, participants remain seated.

\subsubsection{Design novelty}
Our O\&M test design was inspired by the MLMT \cite{chung2018novel}. To overcome the potential dangers associated with physical movement during the test and to eliminate bias caused by fear of an unfamiliar environment, we implemented a torso-steering metaphor for navigation. The direction of movement is controlled by rotating a swivel chair to which a VR controller is attached, allowing participants to visually explore in all directions while moving in a specific direction. The avatar's movement speed within the environment is fixed and intentionally set to a slow pace to prevent cybersickness. Participants hold another controller in their dominant hand, which they can use to start or stop navigation by pressing the trigger at any time.

In previous O\&M tests, the goal was typically to avoid obstacles: participants were instructed to navigate while avoiding collisions, and the number of collisions was used as a metric to evaluate their performance. However, we believe that in virtual reality, collision detection can be easily biased by accidental touches, making the number of collisions an unreliable indicator of obstacle avoidance ability. Therefore, in our test design, we employ an obstacle destruction mechanism to determine whether an obstacle has been detected by the participant. If participants identify an object, they can destroy it by touching it with the virtual hand (represented by the controller in their dominant hand) for 2 seconds, after which the object disappears accompanied by vibration feedback. In this context, the destruction of an obstacle reliably indicates that it has been detected. Consequently, the number of non-destroyed obstacles serves as a more accurate metric for assessing obstacle avoidance ability in our test.

\subsubsection{Virtual environment development}
Ophthalmologists, engineers, and data science researchers collaborated extensively to make corrections and improvements throughout the development of the test \cite{crozet2023virtual}. Preliminary evaluations were conducted with volunteers to identify and rectify any issues in the environment. It took nine months to produce the final version of our virtual environment for this study.

\begin{figure}[htbp]
    \centering
    \begin{subfigure}[b]{0.23\textwidth}
        \centering
        \includegraphics[width=\linewidth, valign=c]{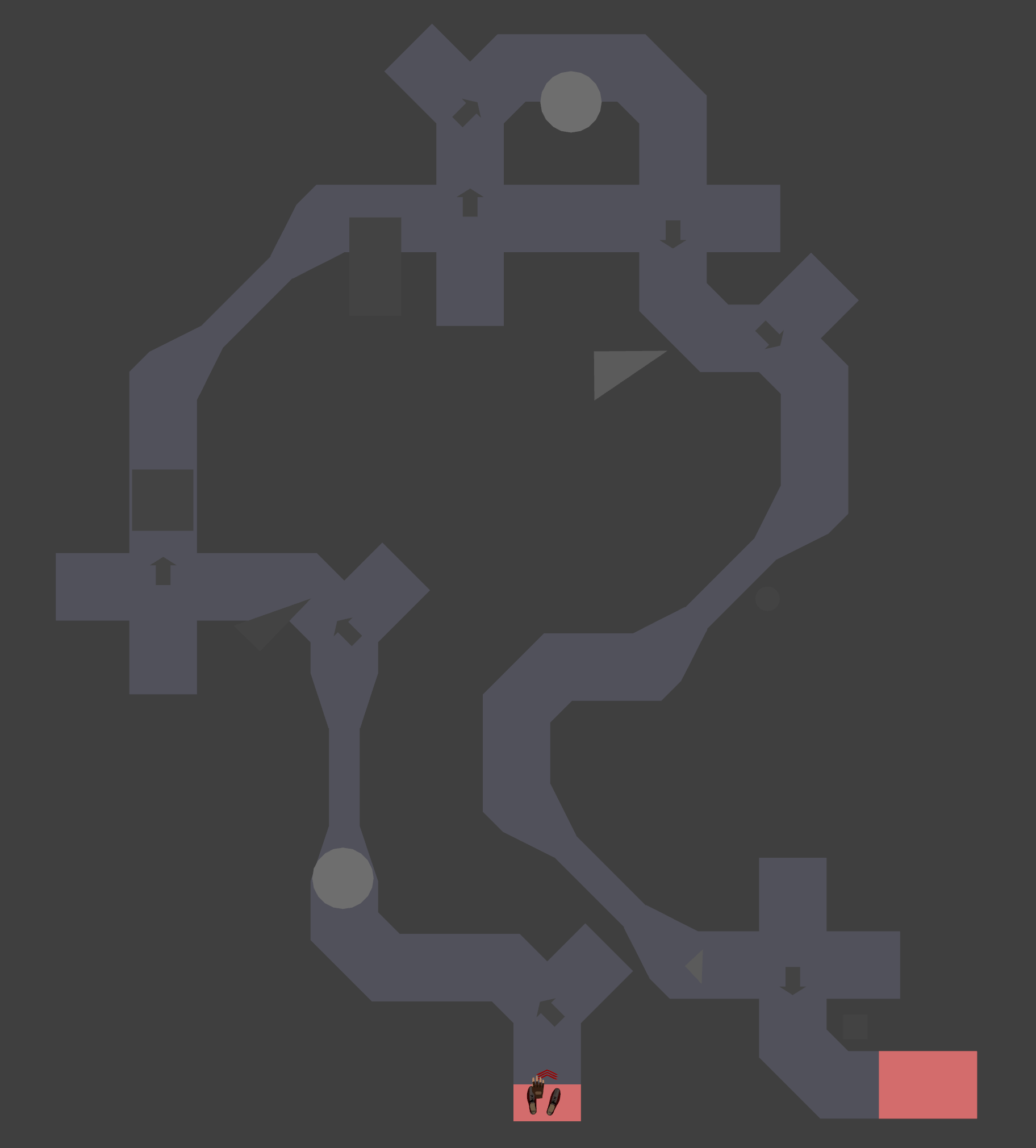}
        \caption{Course A}
        \label{fig:A}
    \end{subfigure}
    \hfill
    \begin{subfigure}[b]{0.23\textwidth}
        \centering
        \includegraphics[width=\linewidth, valign=c]{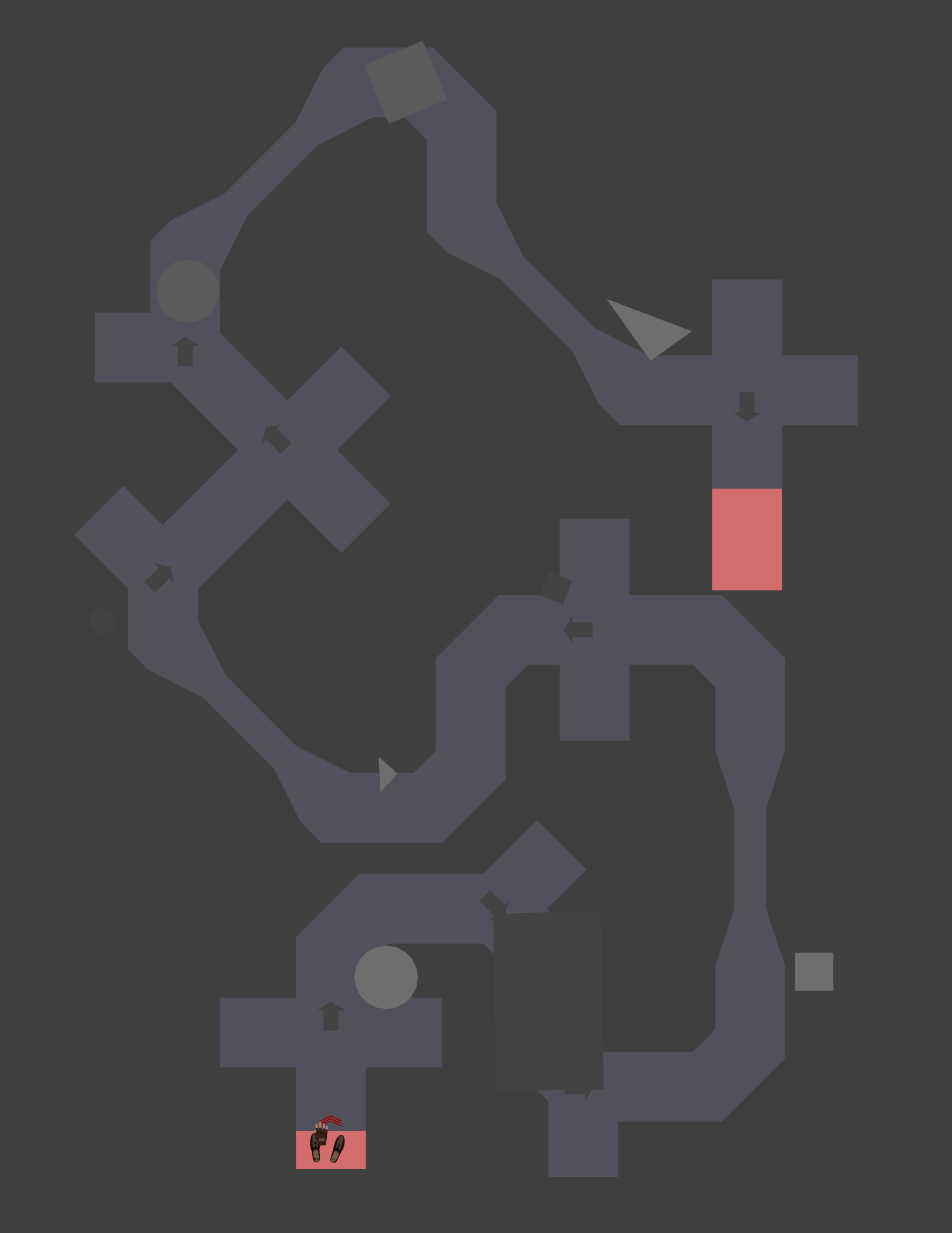}
        \caption{Course B}
        \label{fig:B}
    \end{subfigure}
    \hfill
    \begin{subfigure}[b]{0.23\textwidth}
        \centering
        \includegraphics[width=\linewidth, valign=c]{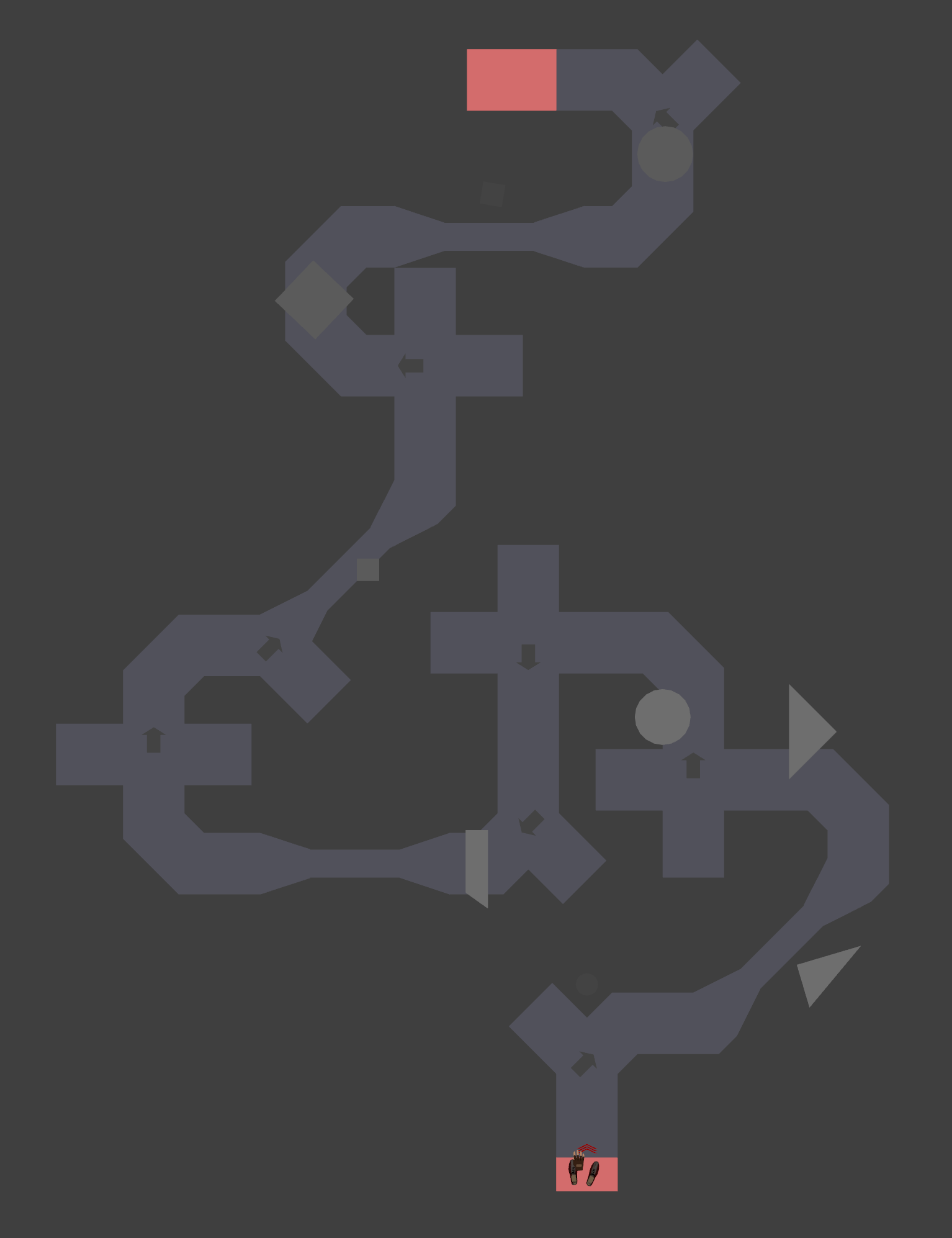}
        \caption{Course C}
        \label{fig:C}
    \end{subfigure}
    \hfill
    \begin{subfigure}[b]{0.23\textwidth}
        \centering
        \includegraphics[width=\linewidth, valign=c]{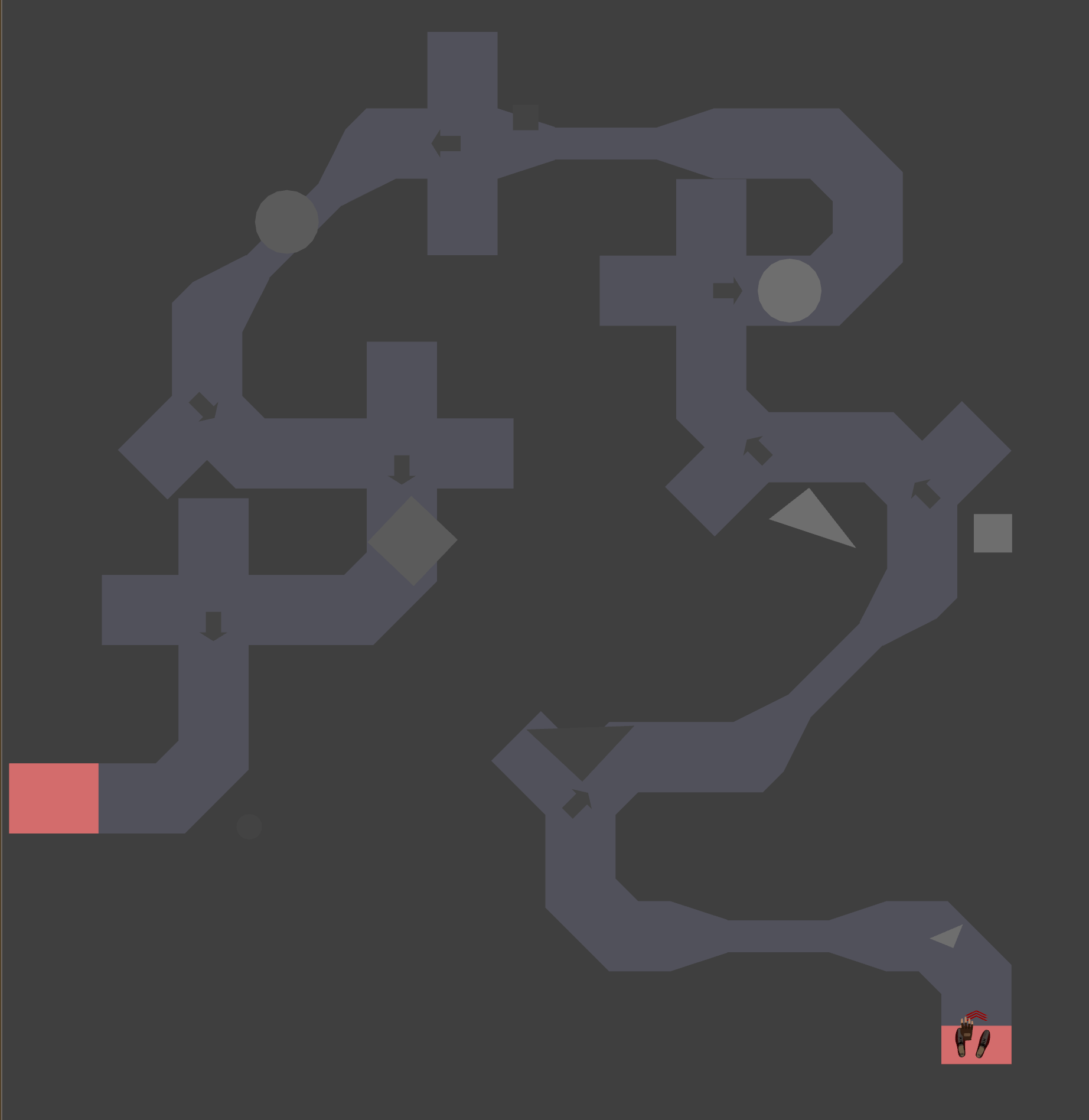}
        \caption{Course D}
        \label{fig:D}
    \end{subfigure}
    
    \vspace{0.5cm}
    
    \begin{subfigure}[b]{0.23\textwidth}
        \centering
        \includegraphics[width=\linewidth, valign=c]{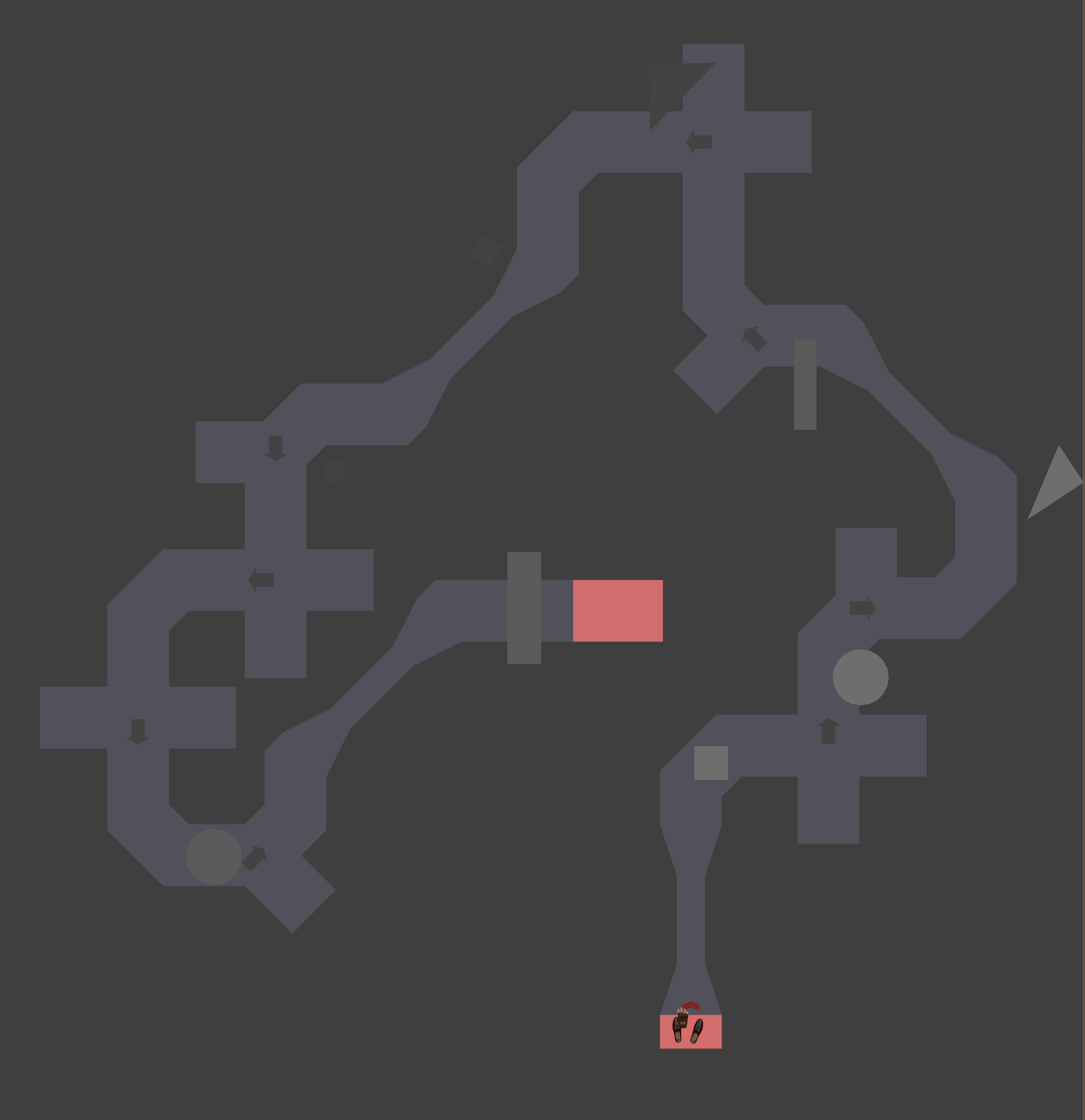}
        \caption{Course E}
        \label{fig:E}
    \end{subfigure}
    \hfill
    \begin{subfigure}[b]{0.23\textwidth}
        \centering
        \includegraphics[width=\linewidth, valign=c]{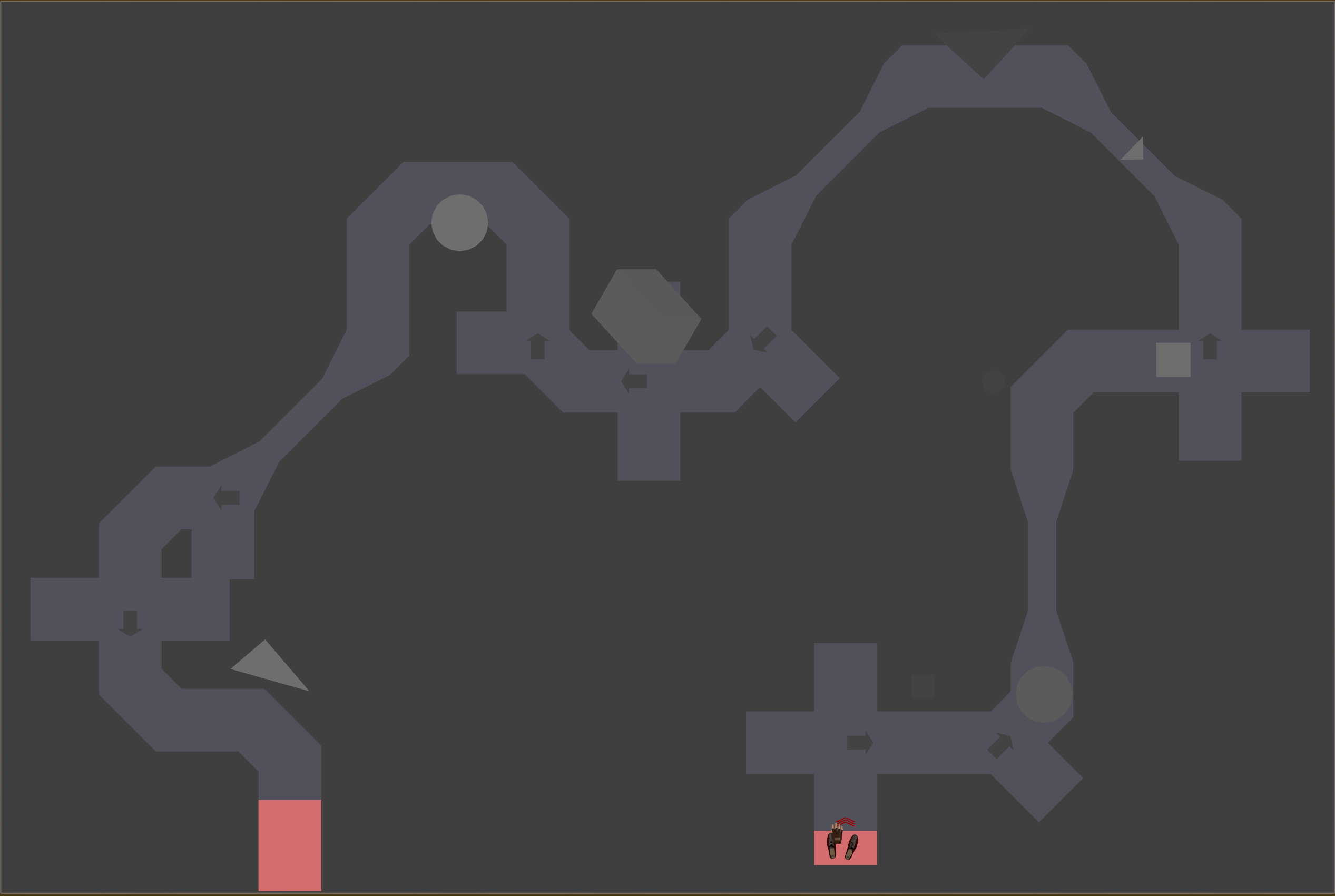}
        \caption{Course F}
        \label{fig:F}
    \end{subfigure}
    \hfill
    \begin{subfigure}[b]{0.23\textwidth}
        \centering
        \includegraphics[width=\linewidth, valign=c]{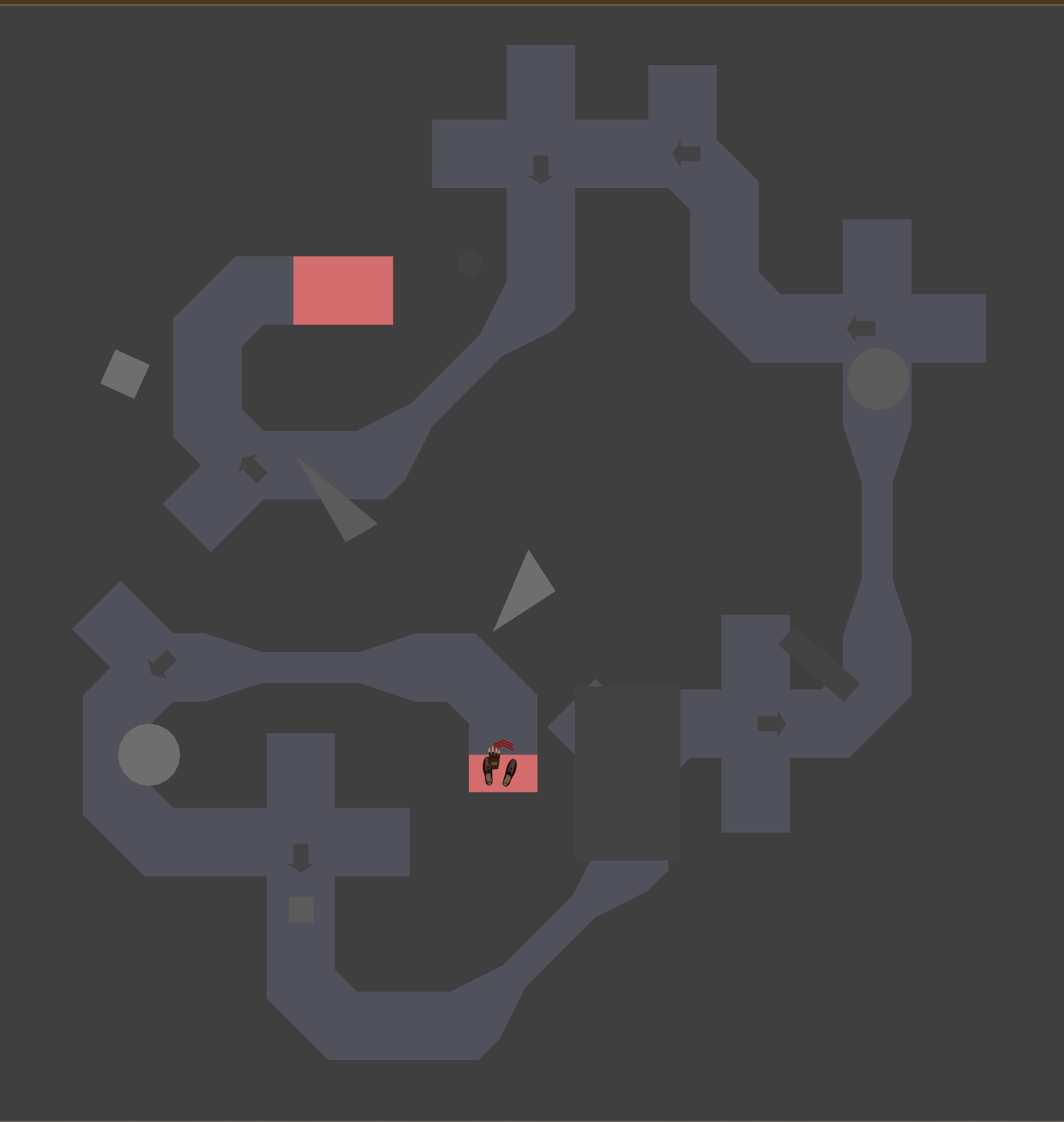}
        \caption{Course G}
        \label{fig:G}
    \end{subfigure}
    \hfill
    \begin{subfigure}[b]{0.23\textwidth}
        \centering
        \includegraphics[width=\linewidth, valign=c]{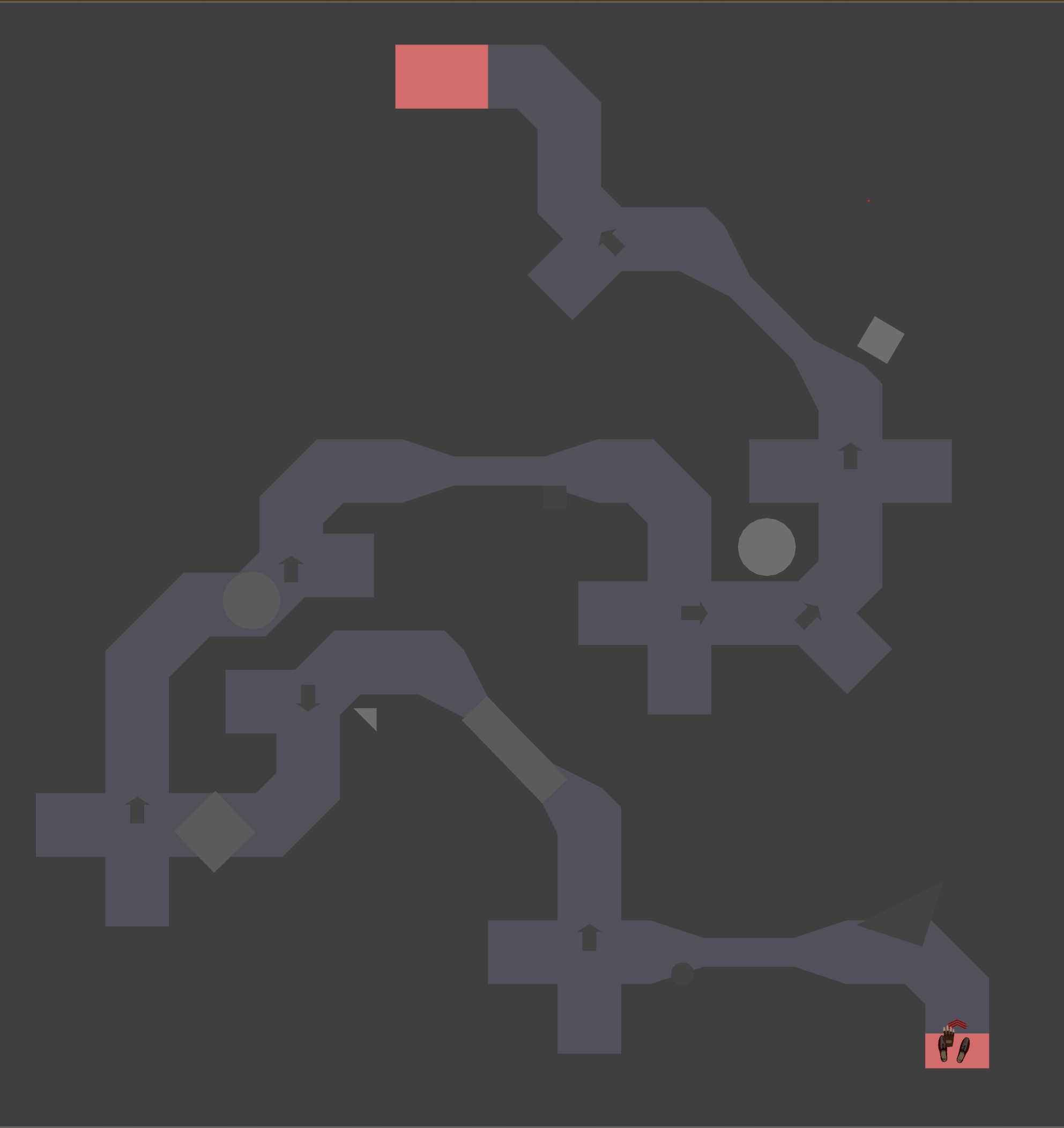}
        \caption{Course H}
        \label{fig:H}
    \end{subfigure}
    \caption{All courses used in this study.}
    \label{fig:all_laby}
\end{figure}

Our VR-S-O\&M test environment includes eight different courses, labeled A to H. Courses A to F are used for evaluation runs, while courses G and H are used for training runs only, and thus their data will not be analyzed. Each course features 4 narrowings, 11 turns, and 9 objects. There are fixed start and end zones in each course, and all courses are of the same length. Aerial views of all the courses used in our study are shown in Figure \ref{fig:all_laby}.

\begin{figure}[H]
    \centering
    \includegraphics[width=0.6\linewidth, valign=c]{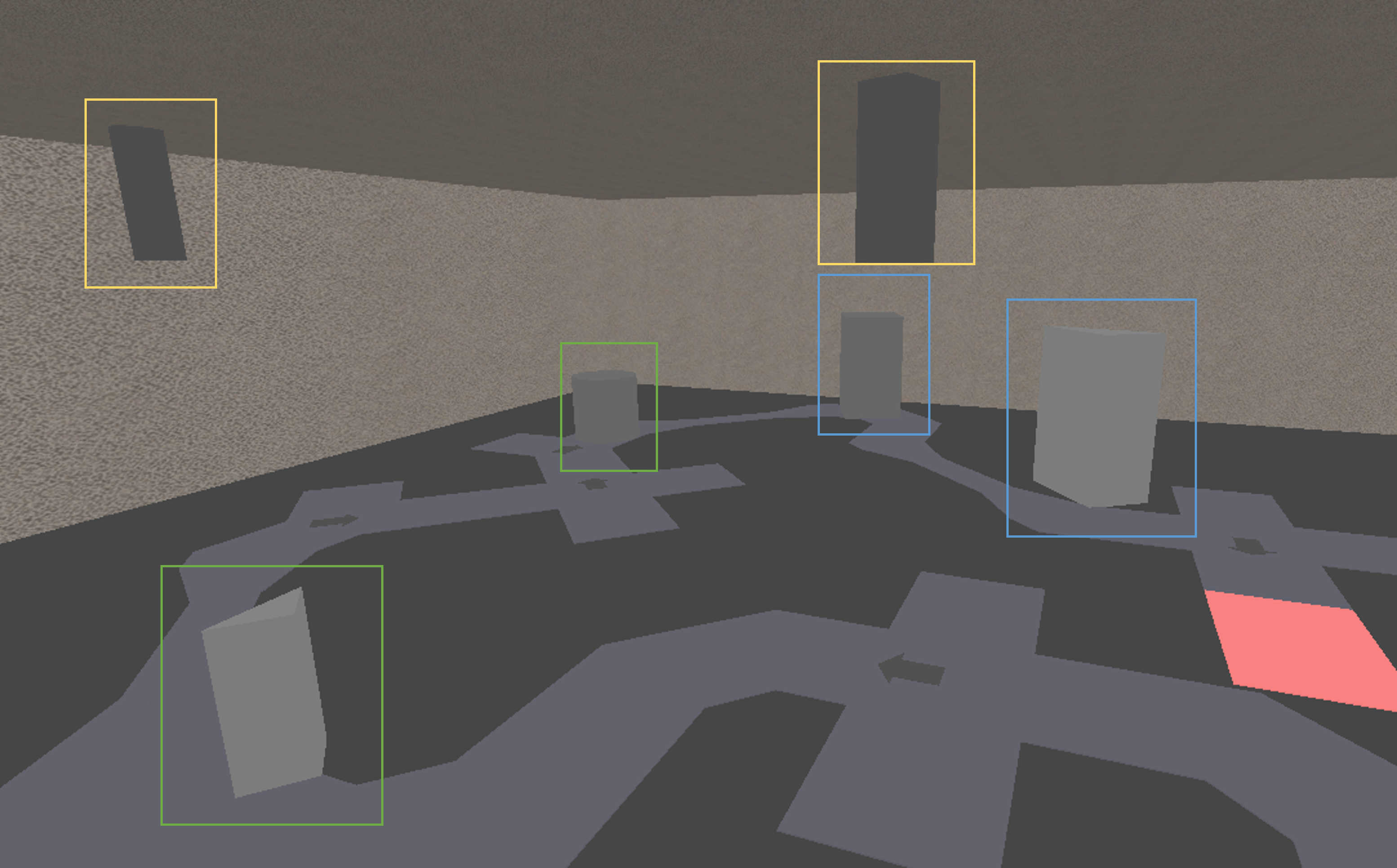}\\
    \caption{Examples of objects in different vertical positions. Green: low; Blue: medium; Yellow: high.}
    \label{fig:obj_vertical}
\end{figure}

\begin{figure}[H]
    \centering
    \includegraphics[width=0.6\linewidth, valign=c]{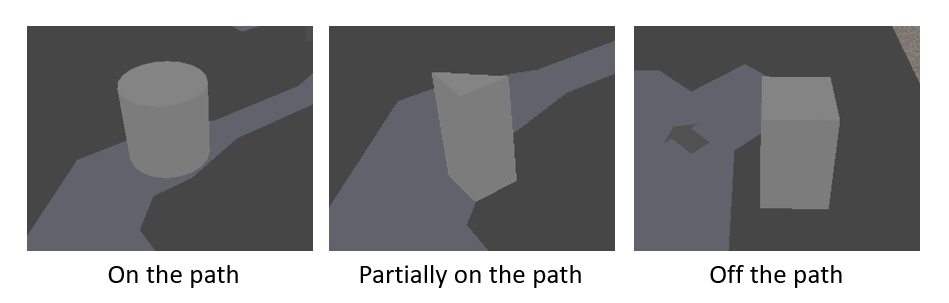}\\
    \caption{Examples of objects in different horizontal positions.}
    \label{fig:obj_horizontal}
\end{figure}

The 9 objects to be destroyed are labeled as follows: cube 0, cube 1, cube 2, cylinder 0, cylinder 1, cylinder 2, pyramid 0, pyramid 1, pyramid 2. Each object has 4 characteristics:
\begin{itemize}
\item Shape: cube, cylinder, pyramid;
\item Vertical position: low, medium and high;
\item Grey level: 83, 111, 134;
\item Horizontal position (relative to the path): on the path, partially on the path, off the path.
\end{itemize}
%Details of each characteristic are provided in Table \ref{tab:char_obj}.
The vertical position is a feature that determines whether the participant is looking down, in front, or up. Thus the objects in the low position (placed on the ground) are designed with a relatively small size, the objects in the medium position (placed on the ground) are designed with a relatively big size, no specific size pattern was set for the objects in the high position (placed above of the avatar's head). The vertical and horizontal position characteristics are shown in Figure \ref{fig:obj_vertical} and Figure \ref{fig:obj_horizontal}.
For the detailed configuration of each object in different courses, refer to the Appendix \ref{app:objs}.

% Additionally, the configuration of the chosen course was provided for each course. Since the courses were constructed using multiple line segments, as shown in Figure \ref{fig:path_line}, the configuration included the coordinates of the endpoints of each line segment and the coordinates of the vertices of each object.
% \begin{figure}[H]
%     \centering
%     \includegraphics[width=0.5\linewidth, valign=c]{figures/path_line.PNG}
%     \caption{Diagram of the course construction.}
%     \label{fig:path_line}
% \end{figure}

% \begin{table}[htbp]
% \centering
% \begin{tabular}{l l l l}
% \hline
% Shape& Height& RGB & Position relative to the path
% \\ \hline
% \begin{tabular}[c]{@{}l@{}}cube\\ cylinder\\ pyramid\end{tabular} & \begin{tabular}[c]{@{}l@{}}high\\ ground\\ low\end{tabular} & \begin{tabular}[c]{@{}l@{}}83\\ 111\\ 134\end{tabular} & \begin{tabular}[c]{@{}l@{}}on the path\\ partially on the path\\ off the path\end{tabular} \\ \hline
% \end{tabular}
% \caption{Options of object characteristic}
% \label{tab:char_obj}
% \end{table}

\begin{figure}[H]
    \centering
    \includegraphics[width=0.8\linewidth, valign=c]{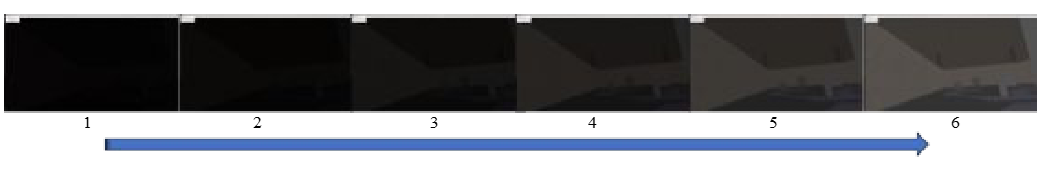}\\
    \caption{Feedback screen of the 6 lighting levels (The display in the VR HMD is different).}
    \label{fig:env_lumi}
\end{figure}

%%%%How to set luminance
We aimed to reproduce controlled variations in lighting, as this is a critical parameter for integration with visually impaired patients, particularly those with hereditary retinal dystrophy. 
Since Unity does not allow direct control of light levels in lux, we measure grey level of the different virtual objects for the 6 lighting levels implemented in our test. 
%Since Unity does not allow direct control of light levels in lux, to ensure that the lighting chosen in Unity was as close as possible to the reality of the MLMT, which is itself calibrated in terms of illumination emitted between 1 and 400 lux, we took measurements at various points along the physical course with a probe measuring luminance in cd/m$^2$. 
Using a colorimetric probe, we then measured the light intensity on the two HMD screens (left and right) for different grey levels displayed. This allowed us to draw calibration curves for the illumination of HMD screens, as well as measuring the light intensity of each virtual object for the 6 lighting levels. The lighting calibration curves and measures are available in the Appendix \ref{app:light}. Figure \ref{fig:sub_view} shows a view of the subject in the virtual reality headset during a test. Figure \ref{fig:env_lumi} shows the 6 lighting levels that can be implemented. 

\begin{figure}[H]
    \centering
    \includegraphics[width=0.6\linewidth, valign=c]{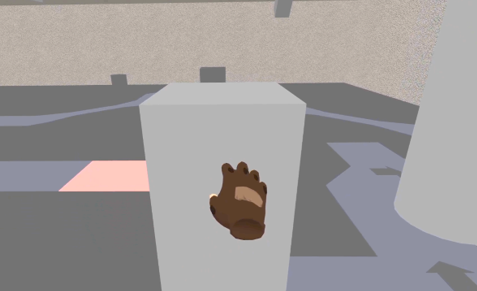}\\
    \caption{Feedback screen of what a subject sees during the virtual mobility test (The display in the VR HMD is different).}
    \label{fig:sub_view}
\end{figure}

\subsection{Experimental protocol}
This study was approved by the local university ethics committee for healthy subjects on November 2022 (IRB number: IORG0011023; Experiment reference number: 16112022). 
%and patients (Comité de Protection des Personnes 24/11/2022, N$^{\circ}$ 2017-A00977-46). 
%A total of 42 healthy subjects were recruited via email from a pre-existing list of individuals deemed suitable for the research. 
Participants were recruited via email from a pre-existing list of individuals deemed suitable for the research. The inclusion criterion was being of legal adult age. The exclusion criteria were as follows:
\begin{itemize}[itemsep=0.05em]
    \item Pregnancy or breastfeeding;
    \item History of epilepsy;
    \item Binocular visual acuity with optical correction less than 8/10;
    \item Lack of depth perception;
    \item Any ophthalmological condition other than the need for corrective lenses.
\end{itemize}
Additionally, participants were required to understand the study rules, be able to hold a controller in their dominant hand, and be capable of turning in a rotating chair using their feet.
% The basic information of subjects are shown in Table \ref{tab:sub_details}.

% \begin{table}[H]
% \centering
% \begin{tabular}{|l|l|}
% \hline
% Number of Subjects                       & 42                                                                                                            \\ \hline
% Age (n=42)                               & 32 (min=18, max=62)                                                                                           \\ \hline
% Gender (n=42)                            & \begin{tabular}[c]{@{}l@{}}Female:19\\ Male:23\end{tabular}                                                   \\ \hline
% Dominant hand                            & \begin{tabular}[c]{@{}l@{}}Right:39\\ Left:\end{tabular}                                                      \\ \hline
% Previous VR experience (n=42)            & \begin{tabular}[c]{@{}l@{}}Yes: 32\\ No: 10\end{tabular}                                                      \\ \hline
% Previous video game experience (n=42)    & \begin{tabular}[c]{@{}l@{}}Yes: 38\\ No: 4\end{tabular}                                                       \\ \hline
% Wearing glasses in the VR headset (n=42) & \begin{tabular}[c]{@{}l@{}}Near vision: 0\\ Far vision: 10\\ Progressive lenses: 4\\ Without: 28\end{tabular} \\ \hline
% \end{tabular}
% \caption{Subjects information}
% \label{tab:sub_details}
% \end{table}

\begin{figure}[!htbp]
    \centering
    \includegraphics[width=0.7\linewidth, valign=c]{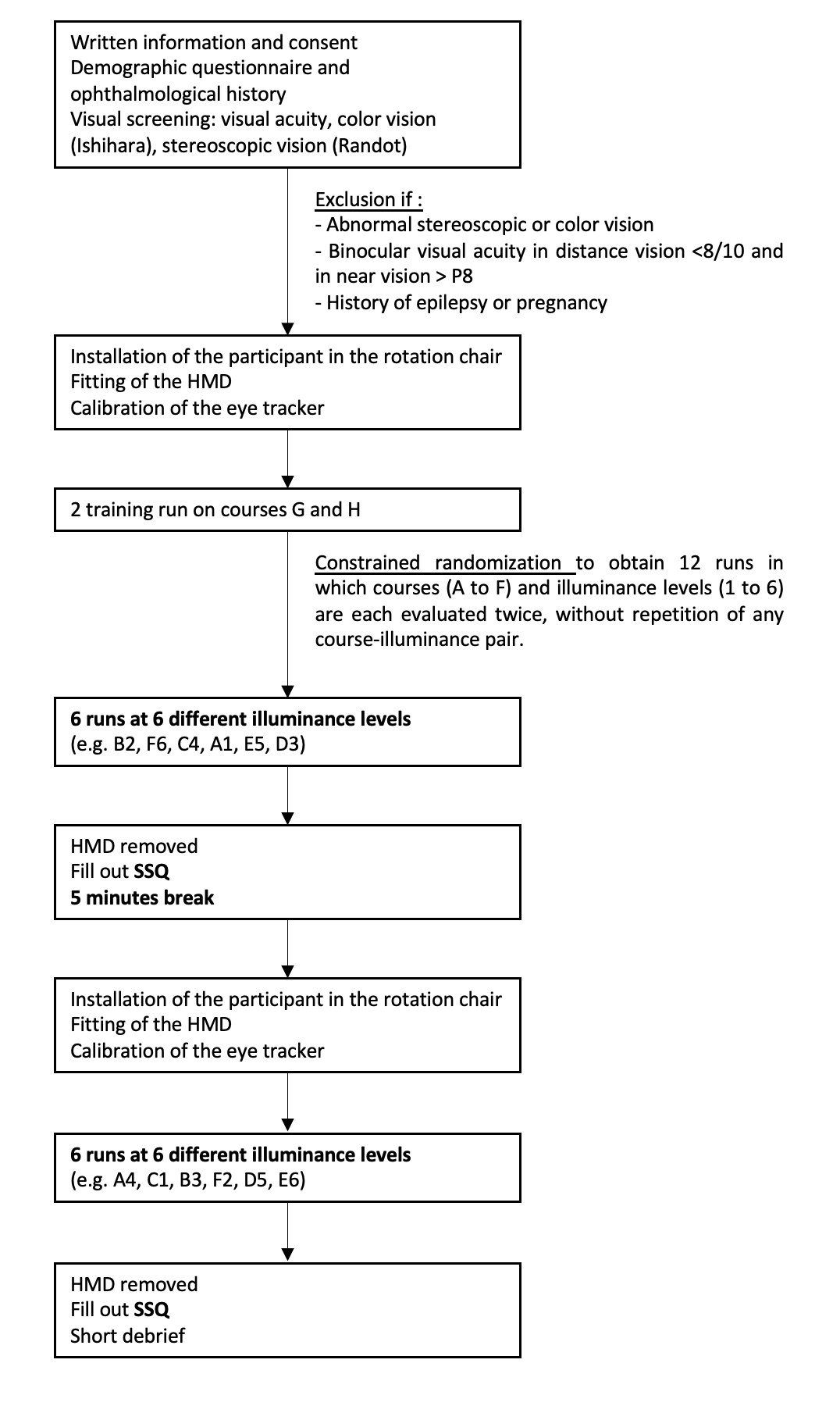}
    \caption{Experimental protocol for healthy subjects.}
    \label{fig:protocol}
\end{figure}

The experimental procedure is shown in Figure \ref{fig:protocol}. Prior to the experiment, participants were provided with oral and written explanations of the experiment. They signed a written consent form, completed a general and ophthalmology history survey, and underwent preliminary vision assessments.

After verifying all exclusion criteria, participants were seated on the swivel chair with the HMD on their head and the controller in their dominant hand. Equipment adjustments and eye tracker calibration were then initiated. Participants were first shown a simple virtual environment consisting of a wooden-tiled floor and floating instructions. After pressing a button, the built-in eye tracker calibration, provided by SRanipal, was performed. Participants adjusted the fitting of the HMD on their head, set the lenses to their individual interpupillary distance (IPD) using a knob on the HMD, and completed five "follow-the-dot" calibration trials. Upon successful calibration, participants pressed a button to begin the experiment.

The experiment began with two training runs using only courses G and H at light level 4. Following the training, participants completed six evaluation courses, took a 5-minute break, and then completed another six evaluation courses. Each course was characterized by a combination of courses (A to F) and light levels (1 to 6). A randomization tool was used to generate these combinations, ensuring that each participant tested each light level twice and each course twice, thereby minimizing the potential for memory effects.

The Simulator Sickness Questionnaire (SSQ) \cite{kennedy1993simulator} was completed by the participant in the middle and at the end of the experiment. 
%The first 9 subjects did not respond to the SSQ. 
%Subject 6, 8, 23, 24 stopped the test because of sickness. 
It contains 16 items to be rated by the participant on a 4-point Likert scale. The items are divided into three subsacles: Disorientation (i.e., dizziness, problems with concentration), Nausea (i.e., general discomfort, nausea) and Oculomotor problems (i.e., eye strain, problems with concentration). Participants were informed that they could ask to stop if they felt uncomfortable at anytime, even if a course was in progress.

At the start of the experiment, participants viewed the route to be completed through the HMD in a closed virtual room. The avatar, with only hands and feet visible, was positioned at the starting point marked by a red plaque on the floor. The timer did not begin until participants started moving forward virtually, allowing them time to analyze the environment before starting. As explained, the goal was to complete the course by destroying all nine objects through touch, following the path, and doing so as quickly as possible.

\subsection{Data collection}
\begin{figure}[H]
    \centering
    \includegraphics[width=0.3\linewidth, valign=c]{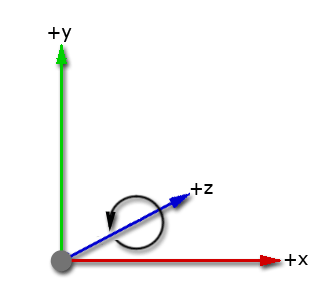}
    \caption{Left-handed coordinate system in Unity.}
    \label{fig:unity_coordinate}
\end{figure}

Using our virtual reality seated orientation and mobility test environment, we collected a preliminary dataset comprising the behaviors of 42 subjects. Since the data were collected using scripts in Unity, which employs a left-handed coordinate system as shown in Figure \ref{fig:unity_coordinate}, all collected data follow this left-handed coordinate system. 
 
The information of the virtual environment can be found in our dataset:
\begin{itemize}[itemsep=0.1em]
    \item Objects placement;
    \item Courses configuration;
\end{itemize}

The HMD and controllers recorded most of the real-time motion data. During the test, the motion data collected included the following:
\begin{itemize}[itemsep=0.1em]
    \item Timestamp;
    \item Localization information of the participant (head, body and hand);
    \item Eye tracking data from Tobii XR SDK.
\end{itemize}

Some user interaction logs during the test were recorded as well:
\begin{itemize}[itemsep=0.1em]
    \item Interaction timestamp;
    \item The initiator of the interaction (e.g. System, User);
    \item The interaction (e.g. destroy, gaze in / out);
    \item The recipient of the interaction (e.g. Object, Path);
    \item Some additional information.
\end{itemize}

Additionally, a file summarizing all runs was created. It contains detailed information, including run order, the labyrinth used, the tested lighting levels, and the experimenter’s notes.

All  details about the data collected and the test configurations in our dataset are explained in Appendix \ref{app:data}.

\subsection{Data Pre-processing}
Data collected from the HMD typically require pre-processing \cite{wu:hal-04429351}. We performed several pre-processing steps for different types of data, as detailed below. The structure of processed data is explained in Appendix \ref{app:data}.

\subsubsection{Head, hand, body data interpolation}
The original recordings of head, hand, and body motion data showed a slight offset from the nominal frequency. To address this, we applied interpolation techniques \cite{illahi2023learning}, specifically linear interpolation for position data and spherical linear interpolation for rotation data \cite{pletinckx1989quaternion}.

The process involved calculating a target timestamp series at the nominal frequency (90 Hz) based on the first and last timestamps in the original data. For each target timestamp $t$, we identified the two closest original timestamps $t_0$, $t_1$ (where $t_0$ is earlier and $t_1$ is later than the target $t$). A ratio was then calculated as follows:
\begin{equation}
    ratio=\frac{t-t_0}{t_1-t_0}
\end{equation}

\begin{figure}[H]
    \centering
    \includegraphics[width=0.6\linewidth, valign=c]{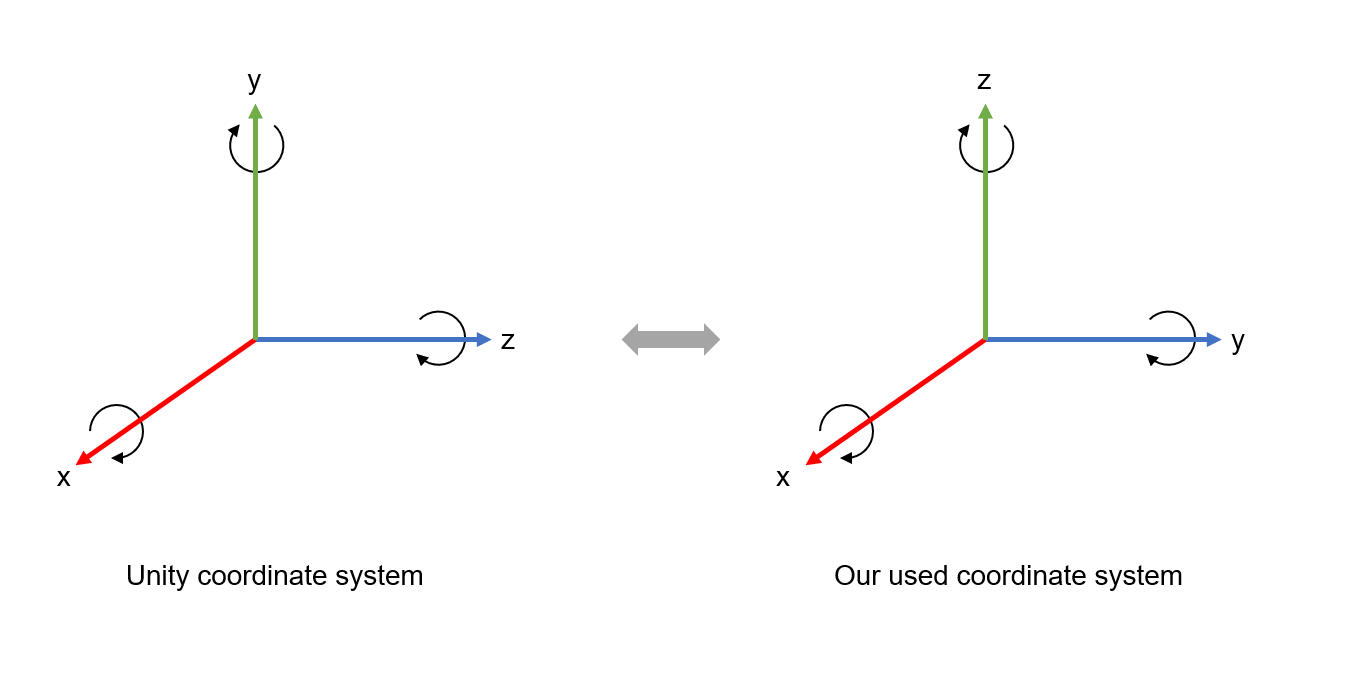}
    \caption{Correspondence of coordinate systems.}
    \label{fig:coor_sys_trans}
\end{figure}

The original data corresponding to these two timestamps were used. For the position data, the interpolated position $\mathbf{p}\in \mathbb{R}^{3}$ is calculated as follows:
\begin{equation}
    \mathbf{p}=\mathbf{p_0}+ratio(\mathbf{p_1}-\mathbf{p_0})
\end{equation}
where $\mathbf{p_0}$ and $\mathbf{p_1}$ are the position data at $t_0$ and $t_1$. For the rotation data $\mathbf{r}\in \mathbb{R}^{3}$, since the original data are in the form of Euler angles and to our knowledge only the quaternion rotation can be interpolated, a transformation from Euler angles to quaternions was performed first. In Unity, rotations represented by Euler angles follow the order of Z-axis, X-axis, and Y-axis. To align this with our coordinate system, as shown in Figure \ref{fig:coor_sys_trans}, the rotation order is "YXZ". The transformation from Euler angles to quaternions was carried out following this order.
The spherical linear interpolation (SLERP) for the quaternion $\mathbf{q}\in \mathbb{R}^{4}$ is calculated as follows:
\begin{equation}
    \mathbf{q}=\frac{sin[(1-ratio)\Omega ]}{sin\Omega}\mathbf{q_0}+\frac{sin(ratio\times \Omega)}{sin\Omega}\mathbf{q_1}
\end{equation}
where $\mathbf{q_0}$ and $\mathbf{q_1}$ are the quaternions at $t_0$ and $t_1$, $\Omega$ is the angle between $\mathbf{q_0}$ and $\mathbf{q_1}$. After this step, all position and rotation data are aligned to a fixed frequency. It is worth noting that, while the interpolated position data still use the left-handed coordinate system as a reference, if future users wish to convert the rotation data to Euler angles, they should follow the same rotation order "YXZ", which refers to the right-handed coordinate system.

\subsubsection{Eye tracking data}
The original eye tracking data were recorded using the Tobii XR SDK and exhibited a time offset relative to the other motion data acquired from SteamVR. To verify the temporal synchronization between eye tracking data and other motion data, we conducted correlation analyses for each course across all subjects. 
\begin{figure}[H]
    \centering
    \includegraphics[width=1\linewidth, valign=c]{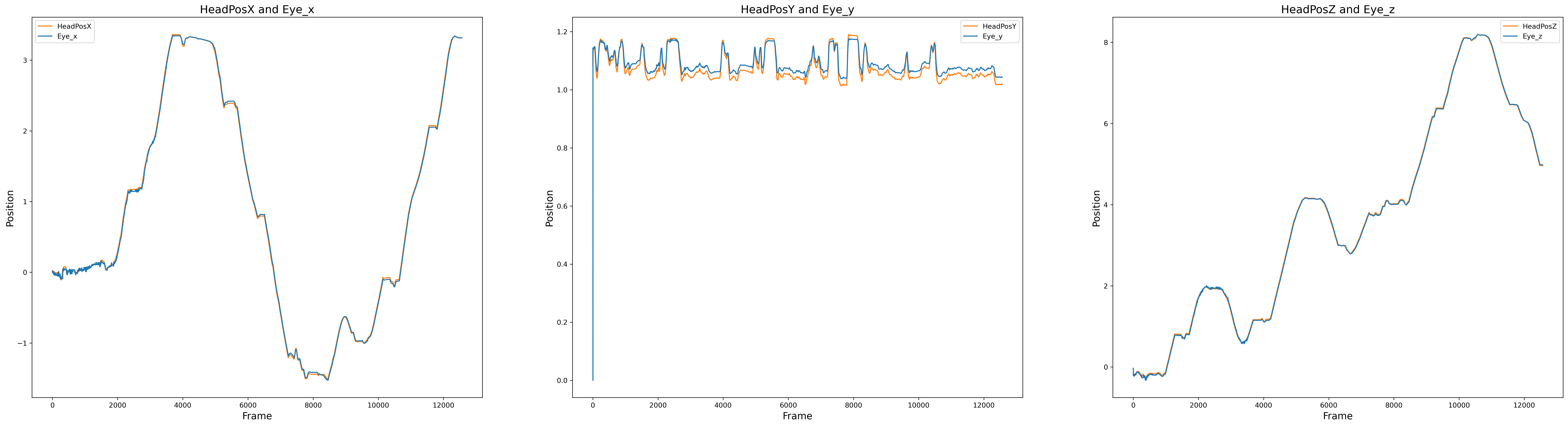}
    \caption{Eye and head position coordinate comparison.}
    \label{fig:eye_head_corr}
\end{figure}
As shown in Figure \ref{fig:eye_head_corr}, we compared the correlation between eye position data and head position data respectively. 93\% of the x position data and 92\% of the z position data (in the left-handed coordinate system) had Pearson correlation coefﬁcients exceeding 0.8 (p value $<$ 0.05), indicating that the time offsets between the eye tracking data and other motion data are consistent. The remaining data that did not show a strong correlation are attributed to invalid eye tracker measurements. Additionally, due to instability in the y position data for eye tracking, it was not included in the temporal synchronization verification.

Therefore, we corrected the timestamps $t_{eye\_corrected}$ for all the eye tracking data $t_{eye}$, without performing interpolation.
\begin{equation}
    t_{eye\_corrected} = t_{eye}-t_{offset}
\end{equation}
where $t_{offset}$ was the difference between the first timestamp of eye tracking data $t_{eye\_0}$ and the first timestamp of motion data $t_{motion\_0}$.

\subsubsection{Synthetized results per run}
To simplify data analysis, we used the events log file and extracted a summarized result for each run. Six key metrics were derived from the collected data.

\textbf{Time duration:} This metric represents the duration (in seconds) from the moment a participant takes the first step until the system ends. To compute this, we traversed the event log file and calculated the time difference between the timestamp of the frame with "system end" and the timestamp of the first frame with "user start."

\textbf{Time before first step:} This metric represents the time interval (in seconds) between the system launch and the participant's first step. We computed this by taking the time difference between the timestamp of the frame with "system launch" and the first frame with "user start."

\textbf{Number of off-path:} This metric represents the number of times a participant deviated from the designated path during a run. We determined this count by identifying the number of "user exit" frames in the log file.

\textbf{Number of missed object:} This metric represents the number of objects that were not destroyed by the participant during a run. To calculate this, we counted the number of "user destroy" frames and subtracted this value from the total number of objects (which is nine in our case).

\textbf{Number of collision:} This metric represents the number of collisions that occurred during a run. We counted the number of "user collide" frames to determine this value.

\textbf{Number of stop} This metric represents the number of times a participant stopped during a run. We counted the occurrences of "user stop" frames in the log file.

\subsubsection{Other issues, data transparency and proposed data usage}
The complexity of this study introduced various issues, many of which are normal and outside of the study design control. We have made corrections for some of these issues. To ensure data transparency, we provide a list of all detected issues, including the corresponding course IDs, the corrections applied, or recommendations for data usage.

\begin{table}[H]
\centering
\resizebox{\columnwidth}{!}{
\begin{tabular}{|l|l|l|}
\hline
\multicolumn{1}{|c|}{Type of issue}                       & \multicolumn{1}{c|}{Subject ID-Run order}                                                                   & \multicolumn{1}{c|}{Solution/Suggestion}                                                                                                                                                       \\ \hline
Interaction file lacks of "system end".                   & \begin{tabular}[c]{@{}l@{}}13-4, 13-6, 13-7, 13-8, 13-9, 13-10, 13-11,\\ 18-10, 27-7, 28-2, 30-1\end{tabular} & Add "system end" manually.                                                                                                                                                                     \\ \hline
Interaction file has "destroy" action after "system end". & 11-10, 26-2                                                                                                   & Remove the "destroy" after "system end".                                                                                                                                                       \\ \hline
Data missing due to subject tolerance.                    & 6-9*, 8-9*, 23-3*, 24-8*                                                                                          & Ignore these courses.                                                                                                                                                                          \\ \hline
Data missing due to technical issue.                      & 17-4, 34-5                                                                                                    & Ignore these courses.                                                                                                                                                                          \\ \hline
Eye tracking data not valid.                              & 36-1*                                                                                                          & \begin{tabular}[c]{@{}l@{}}If eye tracking data is needed for the research, ignore these courses.\\ If eye tracking data is not needed for the research, motion data can be used,\end{tabular} \\ \hline
\end{tabular}}\\
Please note: * indicates this course and all subsequent runs \\
% \captionsetup{singlelinecheck=false, justification=centering}
\caption{Data issue detail}
\end{table}

\subsection{Test performance metrics}
The O\&M tests similar to ours, including the MLMT, VR-O\&M test and its optimization version \cite{chung2018novel, aleman2021virtual, bennett2023optimization}, employ the same scoring system to differentiate subjects with IRDs from normal subjects:
\begin{equation}
    TimeScore=t_{duration}+t_{penalties}
    \label{eq_ts}
\end{equation}
\begin{equation}
    AccuracyScore = \frac{N_{penalties}}{N_{obstacles}}
    \label{eq_as}
\end{equation}
where $t_{penalties}$ was set to 15 seconds added for most types of errors in the MLMT test, including collisions, going off-course, and bypassing tiles, and 30 seconds for redirection errors.

Other O\&M tests use similar principles for performance metrics, primarily based on the time taken to complete the test and the ability to avoid obstacles. For instance, the O\&M test described in \cite{geruschat2012orientation} evaluated the effect of an artificial silicon retina (ASR) implant on mobility by measuring the time taken to walk the course and the number of obstacle contacts. Similarly, the obstacle course used to evaluate a portable collision warning device for individuals with peripheral field loss (PFL) \cite{pundlik2015evaluation} assessed performance based on the percentage of preferred walking speed (PPWS) and the number of collisions.

Since the mechanism for identifying obstacles in our study differs from that used in the MLMT, the proposed scoring system from MLMT is not applicable to our VR-S-O\&M test. Therefore, we rely on basic performance metrics: the time duration and the number of errors. In this context, only "missing an object" is classified as an error. Exits from the path are not studied, as they are very rare in healthy subjects and already lead to an increase in the time taken to complete the path. Additionally, the time taken by participants before their first move is considered, as it reflects the period needed for participants to orient themselves \cite{aleman2021virtual}. By analyzing these metrics in healthy subjects, we aim to assess the effects of environmental factors on functional vision and establish a baseline performance for future research involving patients with visual impairments.

We also analyzed the errors made by healthy participants during the test to gain insights into factors that may lead to overlooking obstacles. This analysis provides valuable understanding of human behavior and can guide further research in developing new scoring metrics.

\section{Results}
\subsection{Subject demographics}
42 healthy participants completed the assessment protocol as described in the Methods section. Their demographic characteristics are detailed in Table \ref{tab:sub_details}. Among those who responded to the recruitment email, all were included in the study except for one individual who, although unaware of it, had no stereoscopic vision. All participants received a financial compensation of 15 euros for their participation. 

\begin{table}[H]
\begin{tabular}{|l|l|}
\hline
\multicolumn{1}{|c|}{Number of subjects}                       & \multicolumn{1}{c|}{42}             \\ \hline
Age (mean, min-max)                                            & 32 (18-62)                          \\ \hline
Gender                                                         & F:19 M:23                           \\ \hline
Dominating hand                                                & R:39 L:3                            \\ \hline
Previous experience of video game                              & Yes:38 No:4                         \\ \hline
Previous experience of virtual reality                         & Yes:32 No:10                        \\ \hline
Wearing optical correction on a daily basis                    & DVL:15, NVL:2, PL:5, No:20          \\ \hline
Wearing spectacles in the HMD                                  & DVL:10, NVL:0, PL:4, No:28          \\ \hline
Wearing contact lenses in the HMD                              & 0                                   \\ \hline
Ophtalmological comorbidities                                  & 0                                   \\ \hline
Refractive surgery                                             & 3                                   \\ \hline
General comorbidities                                          & 2                                   \\ \hline
Stereoscopic vision (Randot test, in arcseconds mean, min-max) & 85 (250-40)                         \\ \hline
Colour vision (Ishihara test plates 2, 10, 22)                 & Normal:42                           \\ \hline
Minimal visual acuity without optical correction               & Binocular DV:10/10, Binocular NV:P5 \\ \hline
\end{tabular}\\
Note: F:Female, M:Male, R:Right, L:Left, DVL:Distance Vision Lens, NVL:Near Vision Lens, PL:Progressive Lens, DV:Distance Vision, NV:Near Vision
\caption{Demographic characteristics of healthy subjects.}
\label{tab:sub_details}
\end{table}

The SSQ scores for the 9 first subjects were not reported due to issues in the questionnaire assignment. 
Four participants (6, 8, 23, 24) stopped the experiment due to cybersickness, resulting in 23 uncompleted courses and 2 courses were missing due to technical problems. This results in a total of 479 evaluation courses completed by the subjects and included in the following statistical analyses. 

%However, 23 courses were missing due to poor tolerance experienced by 4 participants, and 2 courses were missing due to technical issues. 

%A total of 479 evaluation courses were completed by the participants. 

%Incomplete data were excluded from the statistical analyses.

%%%%%%%%%%% list the detail on the appendix %%%%%%%%%%%%
% The dataset is housed in a single repository, which includes a general synthesized file, course configurations, and data for participants 1 through 42. The structure of the data is illustrated in Figure \ref{fig:data_structure}. Raw motion data are recorded in text files, while processed and corrected data are stored in CSV files.
% \begin{figure}[H]
%     \centering
%     \includegraphics[width=1\linewidth, valign=c]{figures/data_structure.PNG}
%     \caption{Data structure.}
%     \label{fig:data_structure}
% \end{figure}

\subsection{VR-S-O\&M subject performance evaluation}
As described in the previous section, we evaluated three metrics to represent participants' performance: the time duration to complete the course, the number of missed objects, and the time before the first step. These metrics reflect participants' speed, ability to identify obstacles, and self-orientation time, respectively. We conducted Kruskal-Wallis analysis across different environmental factor groups to assess their impact on these metrics. The results, shown in Table \ref{tab:Kruskal-Wallis_variable}, indicate significant differences in all three metrics across varying lighting levels, suggesting that illumination affects these variables. However, only the number of missed objects showed a significant difference between course configurations, indicating that while course design affects the difficulty of object destruction, it does not significantly influence test duration or self-orientation time.

\begin{table}[H]
\centering
\resizebox{\columnwidth}{!}{
\begin{tabular}{ccccccc}
\hline
\multirow{3}{*}{Data} & \multicolumn{6}{c}{Environmental Factor}                                                             \\ \cline{2-7} 
                      & \multicolumn{2}{c}{Lighting Level} & \multicolumn{2}{c}{Course}     & \multicolumn{2}{c}{Run Order} \\ \cline{2-7} 
                      & F statistic     & P value           & F statistic & P value        & F statistic      & P value      \\ \hline
Time duration & 19.684           & 0.001 **            & 2.534       & 0.771       & 16.811            & 0.114      \\ \hline
Number of missed object      & 115.853           & 2.369e-23 ****             & 57.886       & 3.320e-11 ****          & 8.030            & 0.711      \\ \hline
Time before first step        & 14.218           & 0.014 *         & 7.368       & 0.195         & 14.920            & 0.186        \\ \hline

\end{tabular}}\\
*p\textless0.05, **p\textless0.01, ***p\textless0.001, ****p\textless0.0001.\\
\captionsetup{singlelinecheck=false, justification=centering}
\caption{Kruskal-Wallis analysis of variables across different groups}
\label{tab:Kruskal-Wallis_variable}
\end{table}

\begin{figure}[!htbp]
    \centering
    \begin{subfigure}[b]{0.3\textwidth}
        \centering
        \includegraphics[width=\linewidth, valign=c]{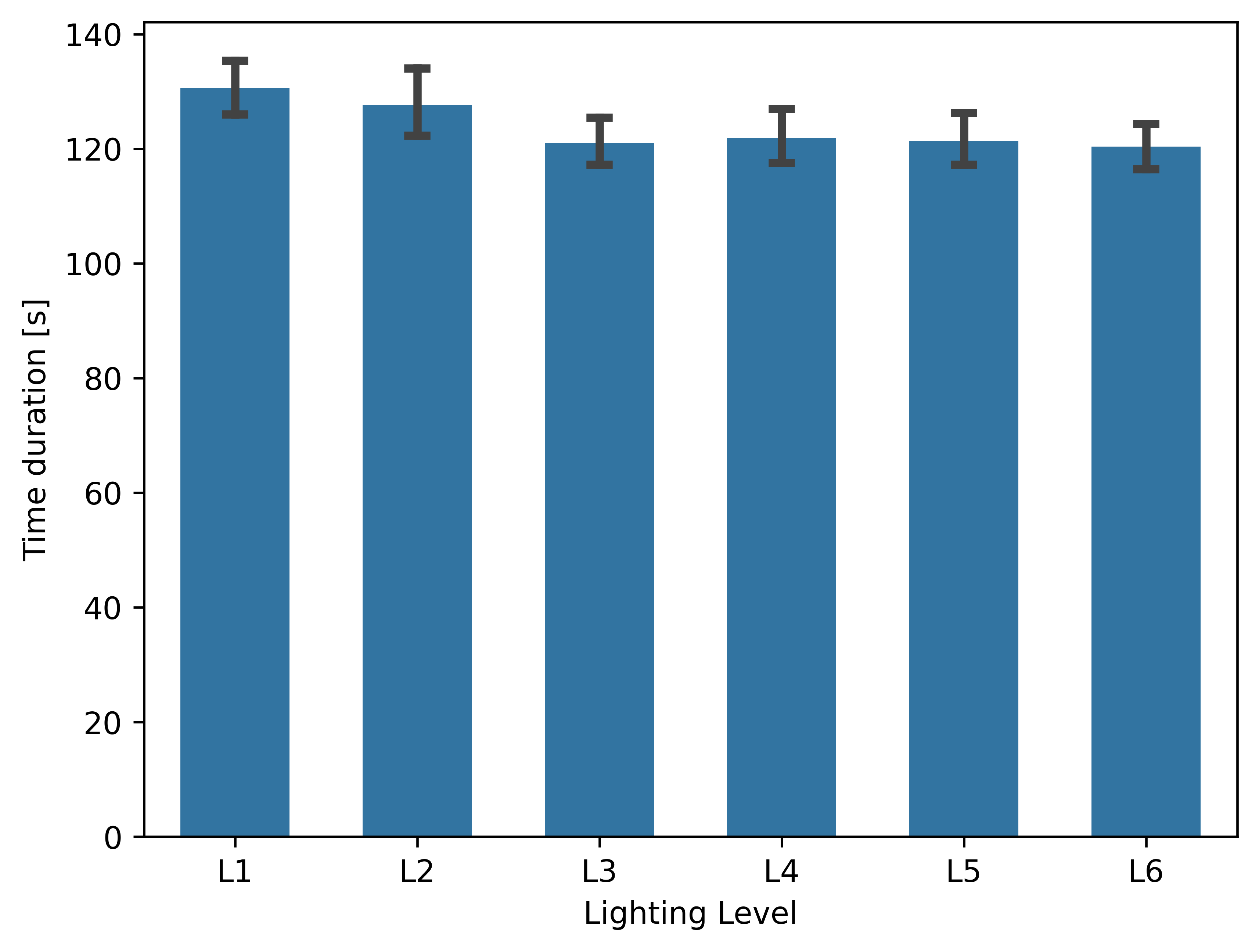}
        \caption{Time duration}
        \label{fig:lumi_timedur}
    \end{subfigure}
    \hfill
    \begin{subfigure}[b]{0.3\textwidth}
        \centering
        \includegraphics[width=\linewidth, valign=c]{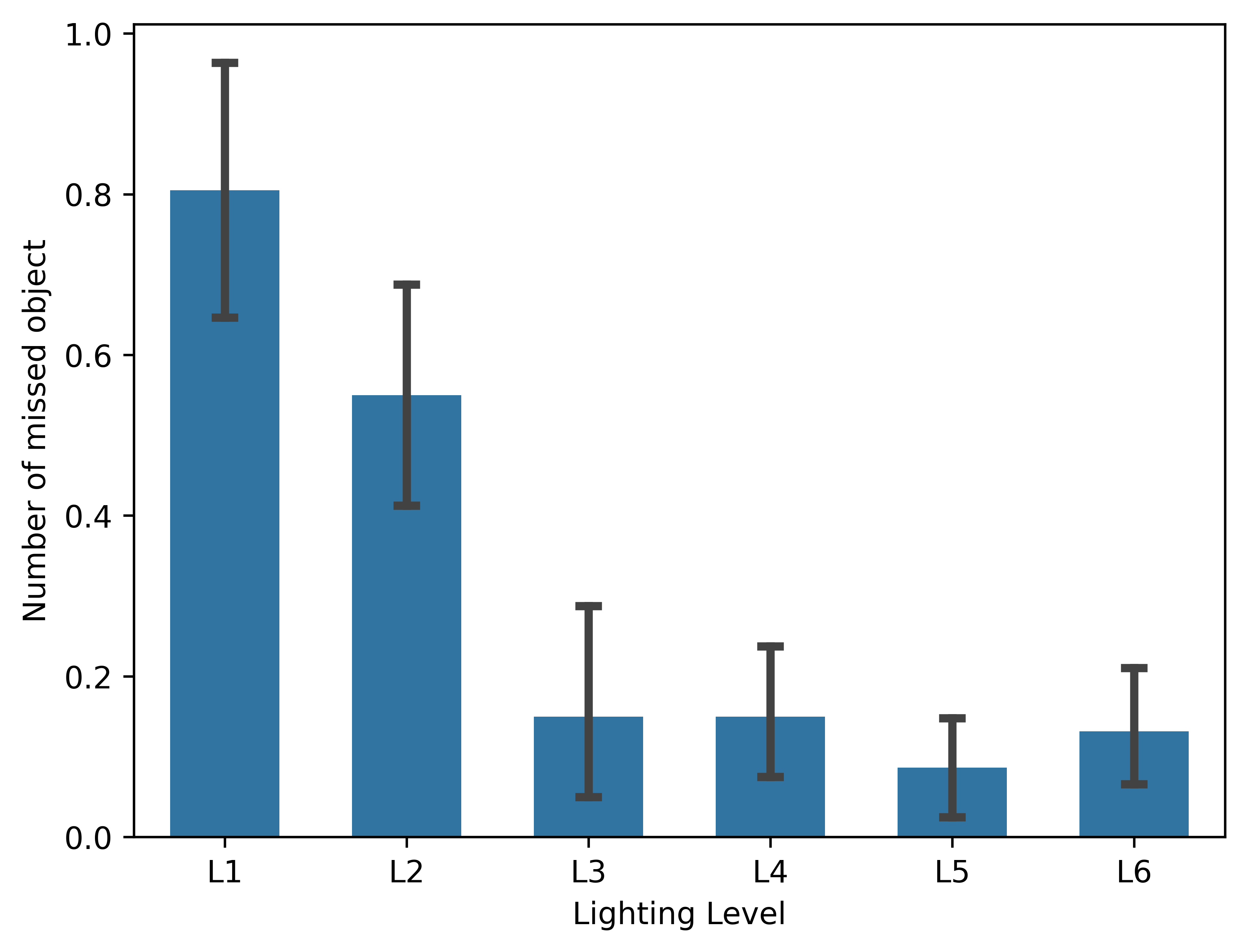}
        \caption{Number of missed object}
        \label{fig:lumi_nberr}
    \end{subfigure}
    \hfill
    \begin{subfigure}[b]{0.3\textwidth}
        \centering
        \includegraphics[width=\linewidth, valign=c]{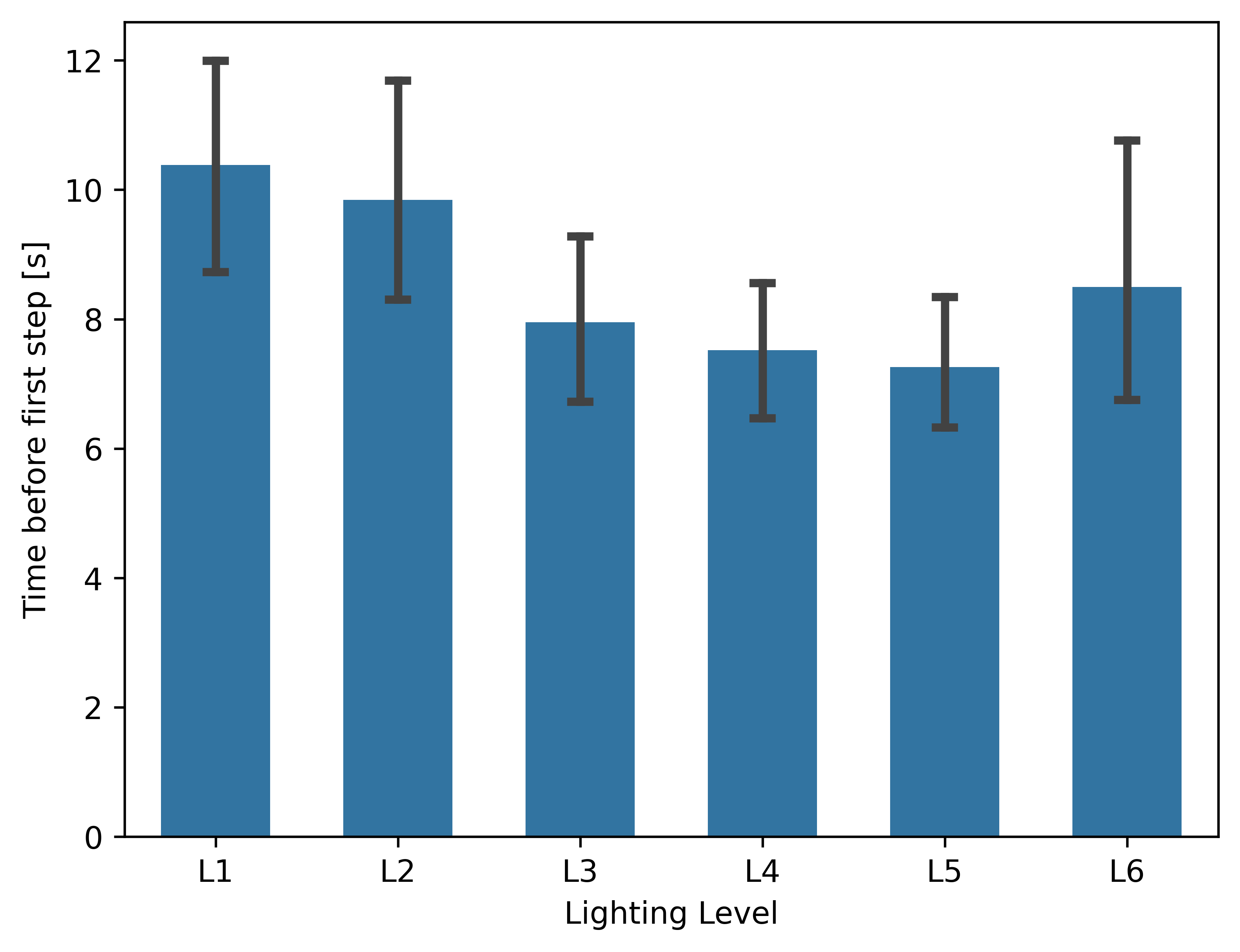}
        \caption{Time before the first step}
        \label{fig:lumi_time_before}
    \end{subfigure}\\
    \caption{Metrics as function of the lighting level of the environment, from L1 (low lighting) to L6 (high lighting).}
    \label{fig:lumi_variables}
\end{figure}

\begin{figure}[!htbp]
    \centering
    \begin{subfigure}[b]{0.3\textwidth}
        \centering
        \includegraphics[width=\linewidth, valign=c]{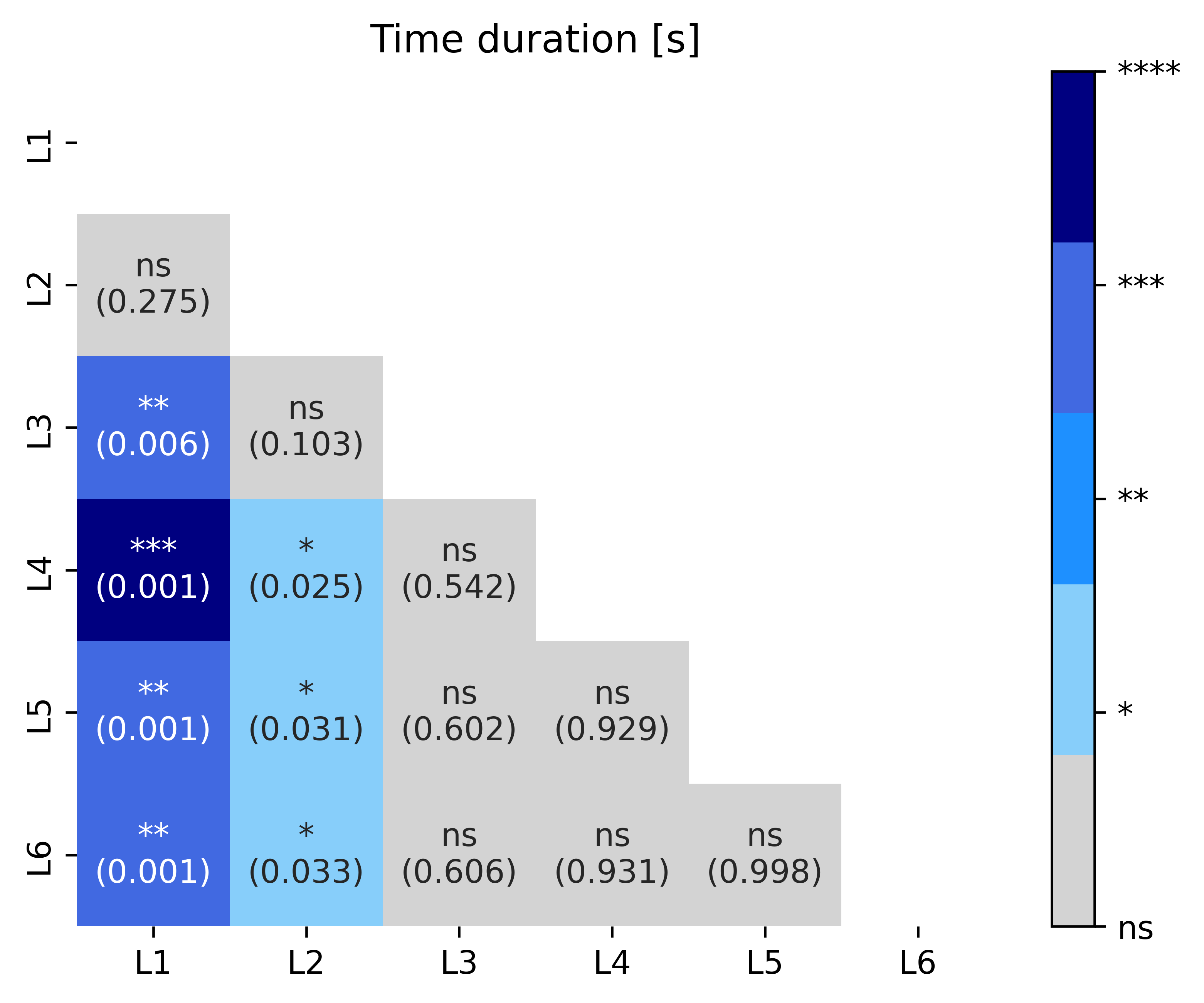}
        \caption{Time duration}
        \label{fig:dunn_lumi_timedur}
    \end{subfigure}
    \hfill
    \begin{subfigure}[b]{0.3\textwidth}
        \centering
        \includegraphics[width=\linewidth, valign=c]{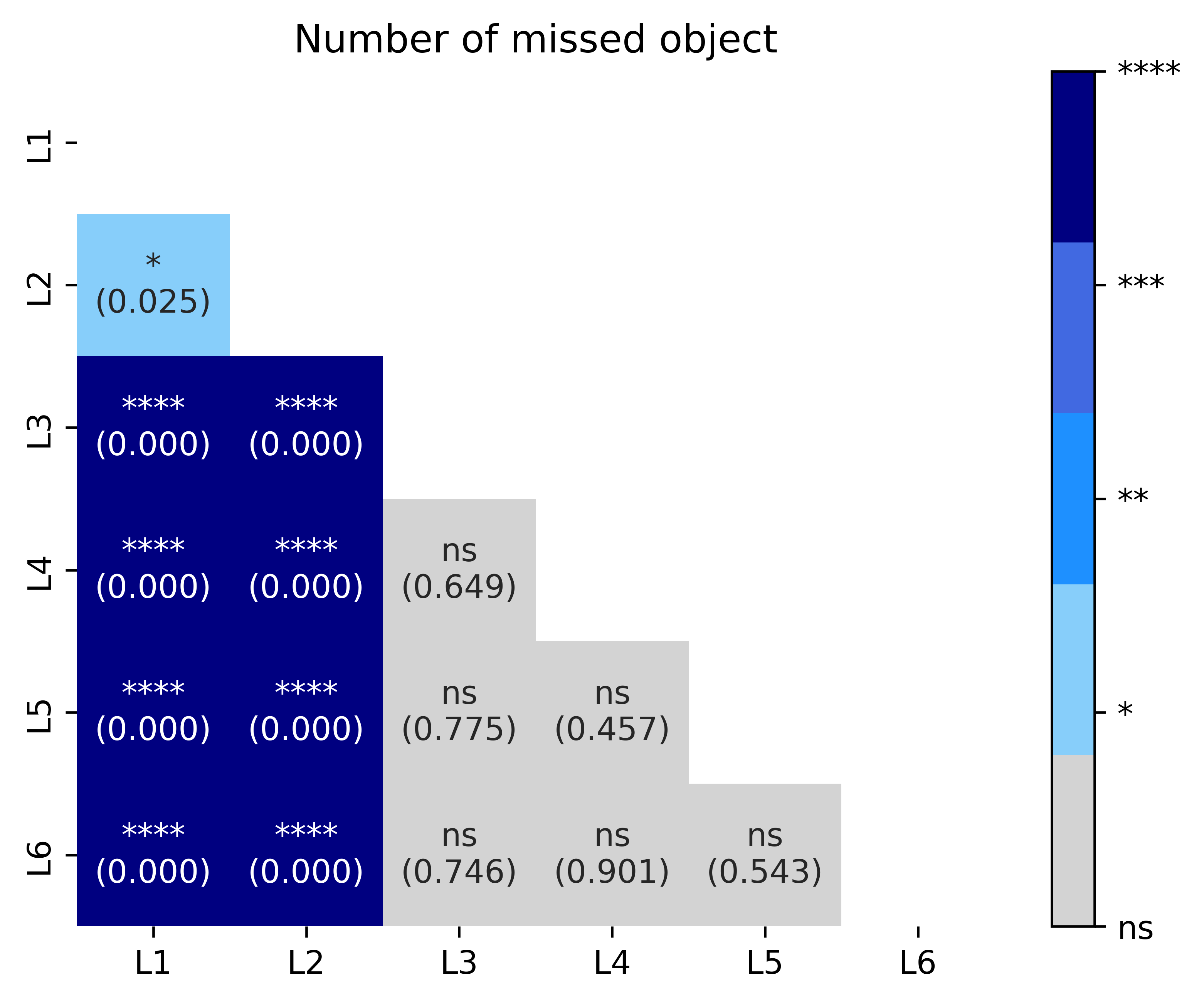}
        \caption{Number of missed object}
        \label{fig:dunn_lumi_nberr}
    \end{subfigure}
    \hfill
    \begin{subfigure}[b]{0.3\textwidth}
        \centering
        \includegraphics[width=\linewidth, valign=c]{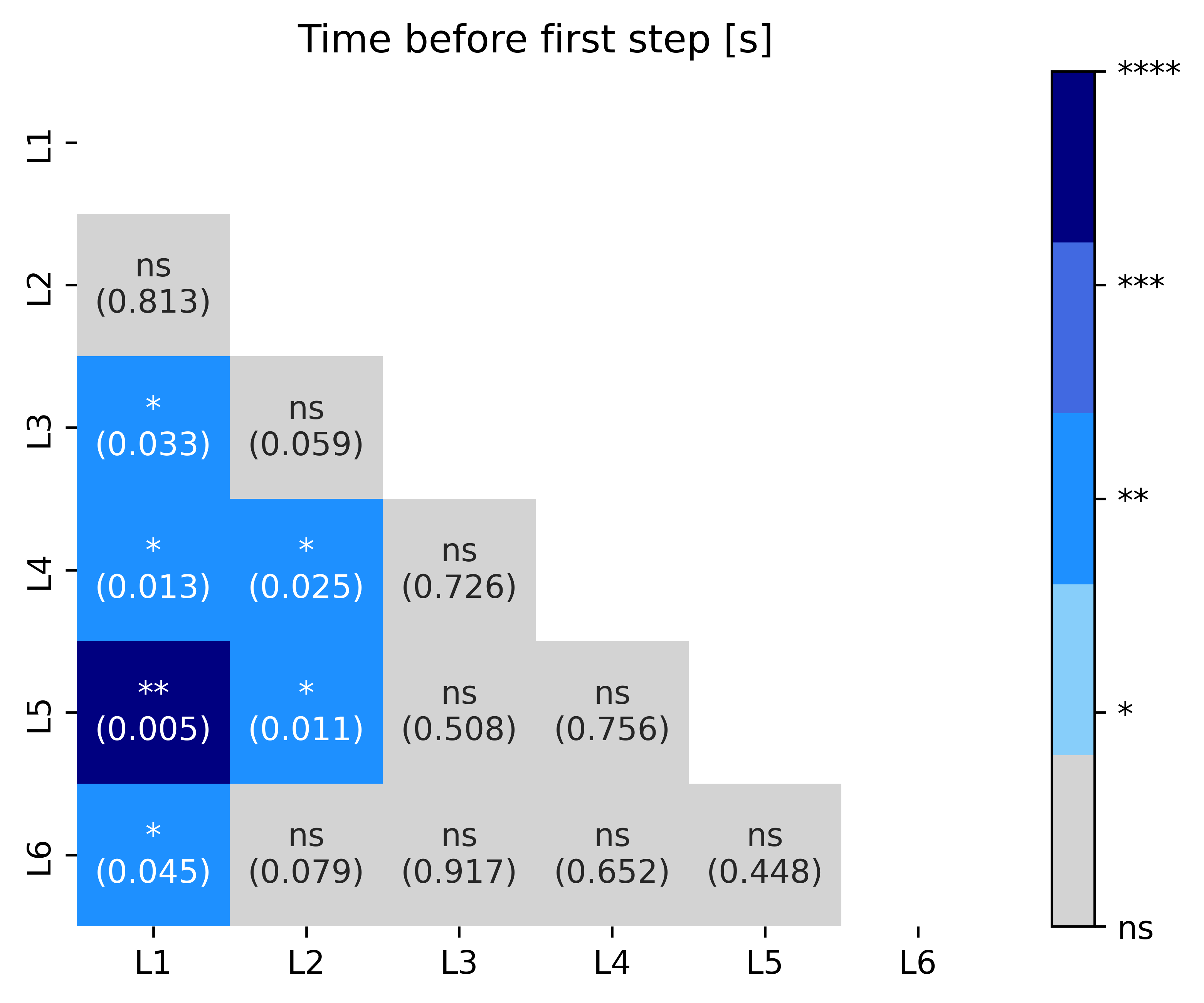}
        \caption{Time before the first step}
        \label{fig:dunn_lumi_time_before}
    \end{subfigure}\\
    \caption{Dunn post-hoc test p-value comparison with lighting levels from L1 (low lighting) to L6 (high lighting).}
    \label{fig:dunn_lumi_variables}
\end{figure}

Figures \ref{fig:lumi_variables}\subref{fig:lumi_timedur}, \ref{fig:lumi_variables}\subref{fig:lumi_nberr}, \ref{fig:lumi_variables}\subref{fig:lumi_time_before} illustrate the impact of lighting levels on the time duration, the number of missed objects, and the time before the first step, respectively. For the first three lighting levels, as the environment becomes brighter, performance metrics improve significantly: participants complete the course more quickly, miss fewer objects, and take less time to start. Beyond the lighting level L3, a threshold effect is observed, where performance improvements plateau and become relatively constant. This effect is further supported by the Dunn test results shown in Figure \ref{fig:dunn_lumi_variables}.

\begin{figure}[H]
    \centering
    \begin{subfigure}[b]{0.3\textwidth}
        \centering
        \includegraphics[width=\linewidth, valign=c]{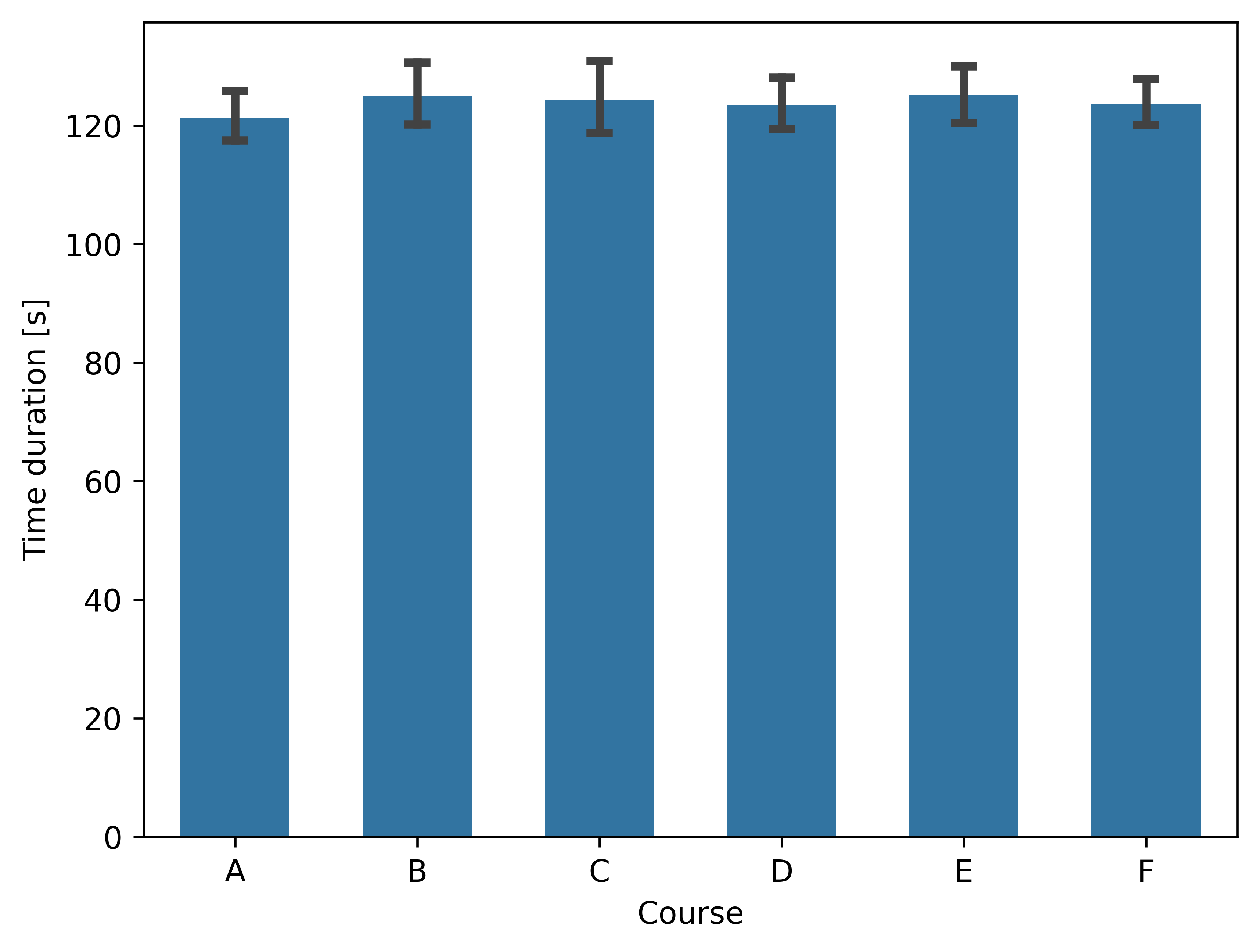}
        \caption{Time duration}
        \label{fig:laby_time}
    \end{subfigure}
    \hfill
    \begin{subfigure}[b]{0.3\textwidth}
        \centering
        \includegraphics[width=\linewidth, valign=c]{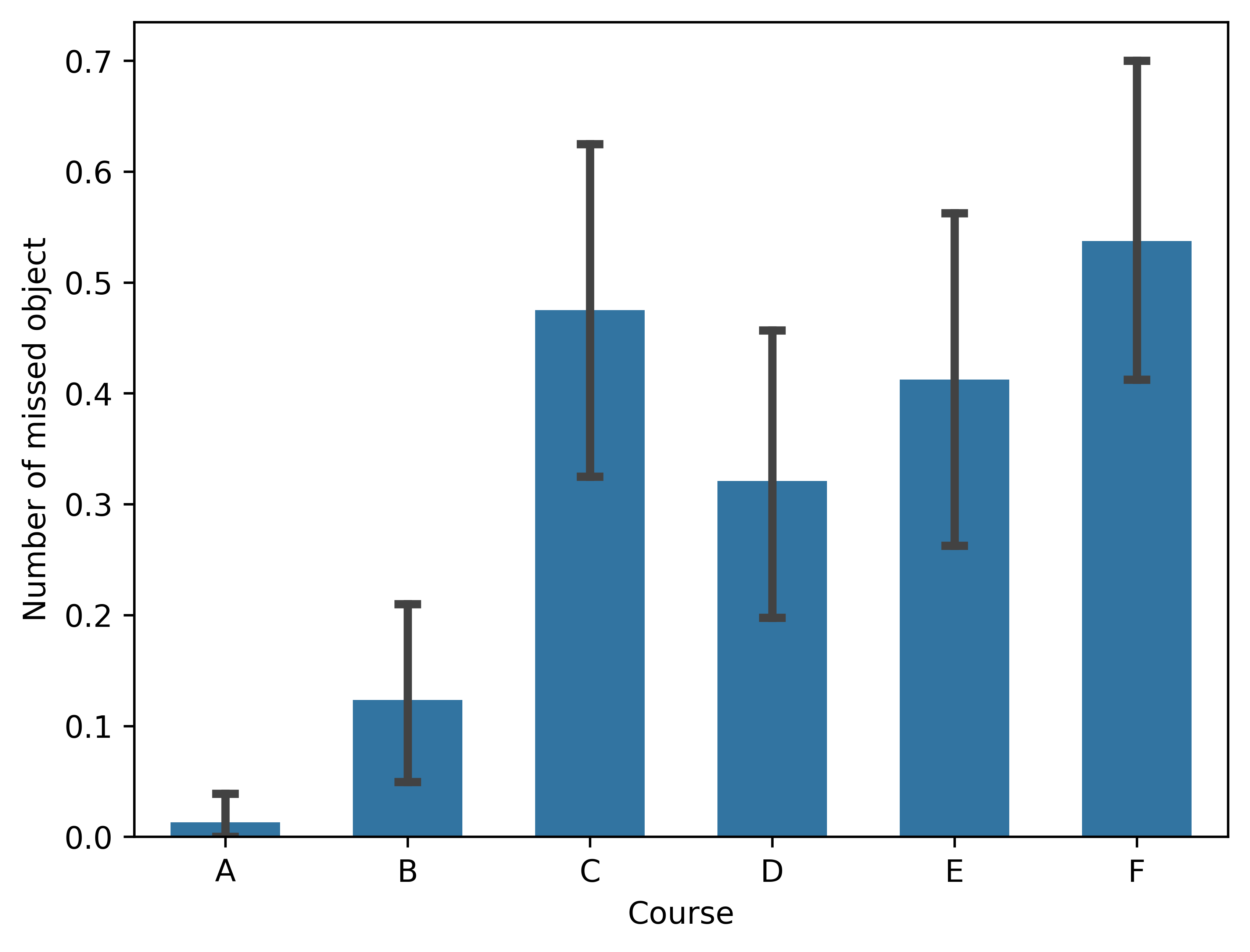}
        \caption{Number of missed object}
        \label{fig:laby_err}
    \end{subfigure}
    \hfill
    \begin{subfigure}[b]{0.3\textwidth}
        \centering
        \includegraphics[width=\linewidth, valign=c]{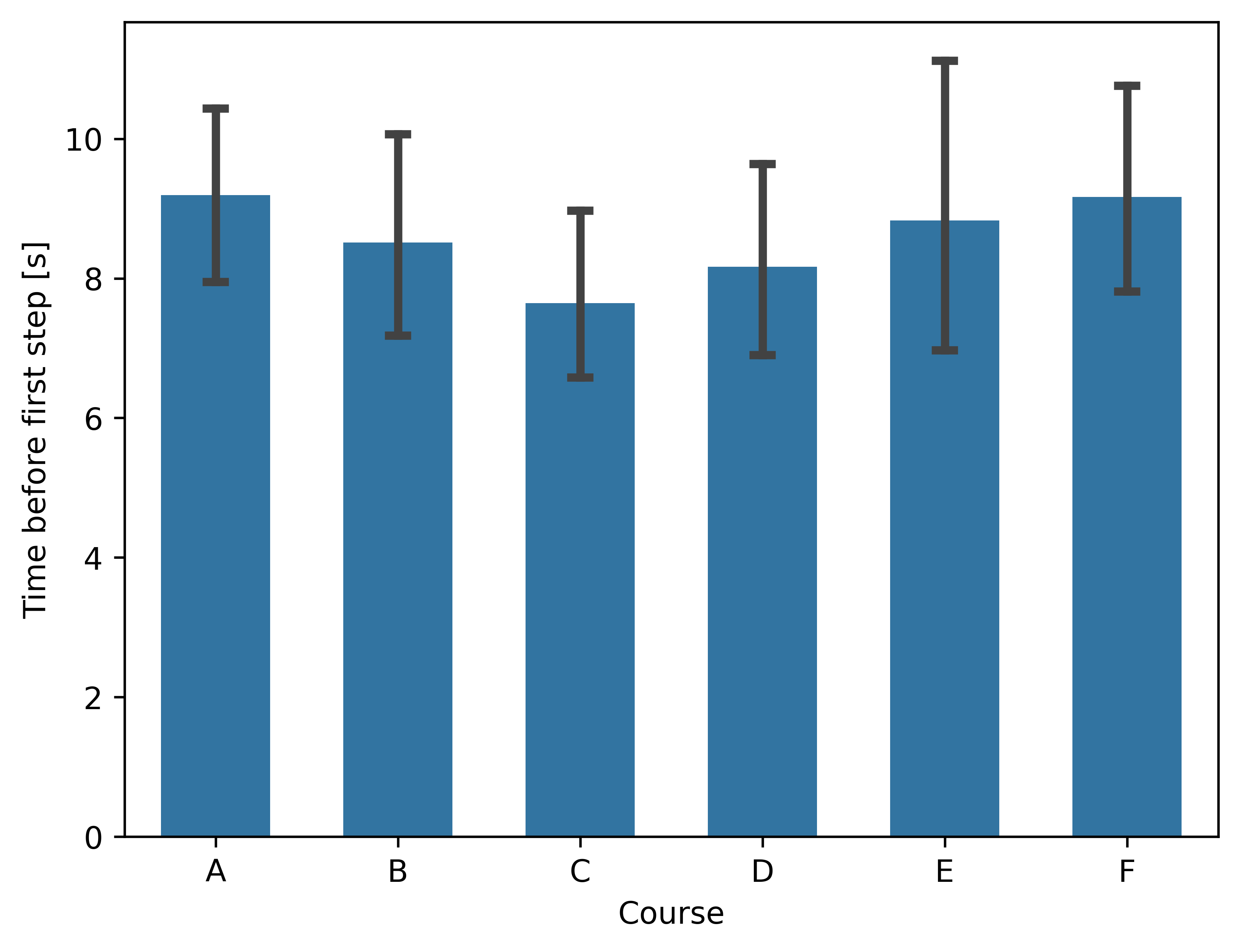}
        \caption{Time before the first step}
        \label{fig:laby_tbfs}
    \end{subfigure}
    \caption{Metrics as function of the course.}
    \label{fig:laby_variables}
\end{figure}

\begin{figure}[H]
    \centering
    \includegraphics[width=0.3\linewidth, valign=c]{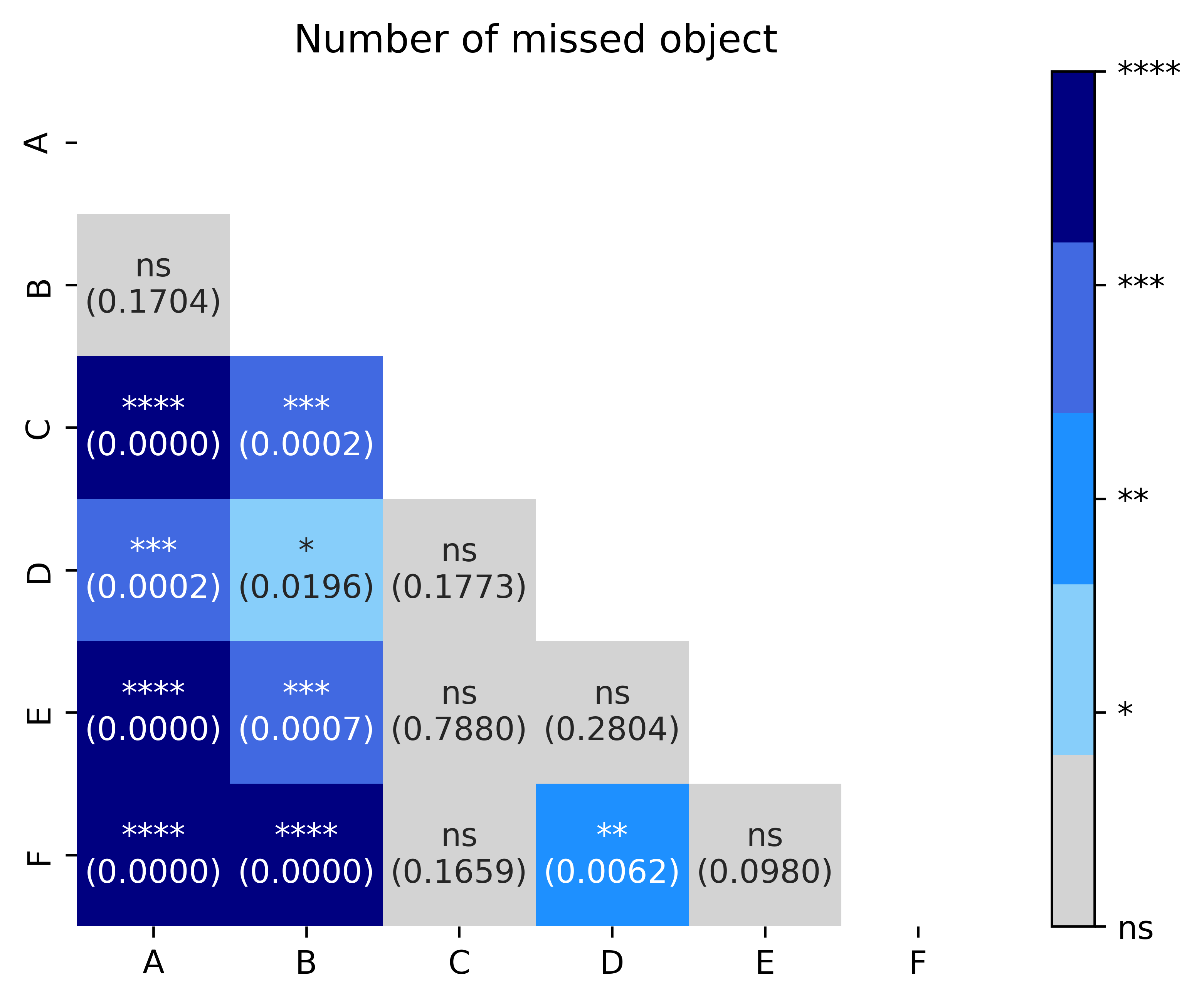}
    \caption{Dunn post-hoc test p-value comparison with courses from A to F}
    \label{fig:dunn_laby_err}
\end{figure}

Figures \ref{fig:laby_variables}\subref{fig:laby_time}, \ref{fig:laby_variables}\subref{fig:laby_err}, \ref{fig:laby_variables}\subref{fig:laby_tbfs} show the comparability among the 6 courses. According to Table \ref{tab:Kruskal-Wallis_variable}, there is no significant difference between courses in terms of time duration and time before the first step. However, there is a significant difference in the number of missed objects. The results, shown in Figure \ref{fig:dunn_laby_err}, indicate that courses A and B resulted in fewer errors, course D had a relatively higher error rate, and courses C, E, and F led to the highest number of errors among healthy subjects.
% Specifically, courses A and B result in fewer errors, while courses C, D, E, and F lead to more errors among healthy subjects.
% % (TODO : change the figure of Number of missed object with ***). 
This indicates that despite our efforts to standardize the course configurations, such as maintaining equal length, the difficulty levels of the courses differ. This discrepancy suggests that certain configurations of the course may contribute to a higher error rate. This information should be considered for future work in order to compare courses with the same level of difficulty. 
%, offering an intriguing research question for our future work.

\begin{figure}[H]
    \centering
    \begin{subfigure}[b]{0.3\textwidth}
        \centering
        \includegraphics[width=\linewidth, valign=c]{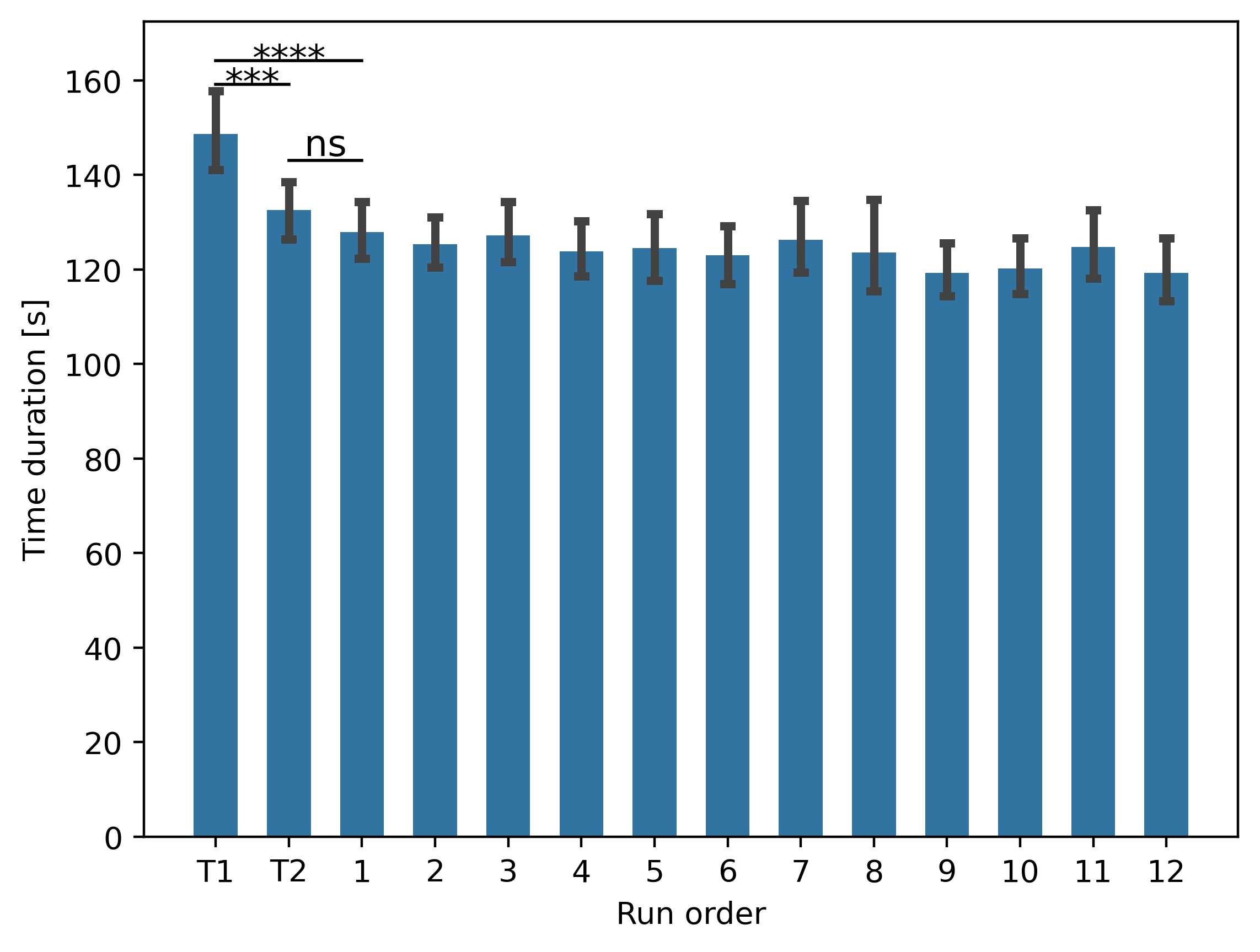}
        \caption{Time duration}
        \label{fig:order_time}
    \end{subfigure}
    \hfill
    \begin{subfigure}[b]{0.3\textwidth}
        \centering
        \includegraphics[width=\linewidth, valign=c]{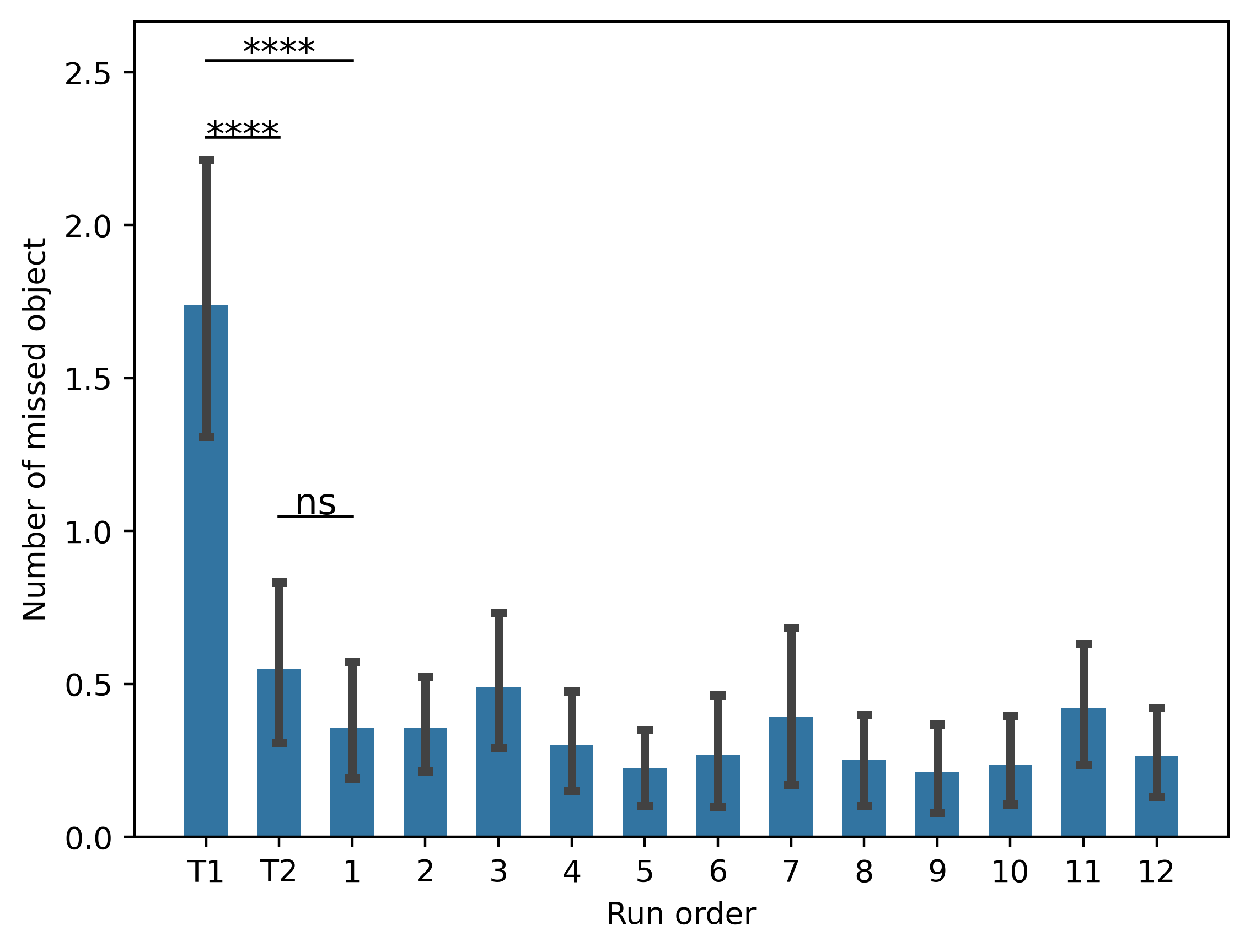}
        \caption{Number of missed object}
        \label{fig:order_err}
    \end{subfigure}
    \hfill
    \begin{subfigure}[b]{0.3\textwidth}
        \centering
        \includegraphics[width=\linewidth, valign=c]{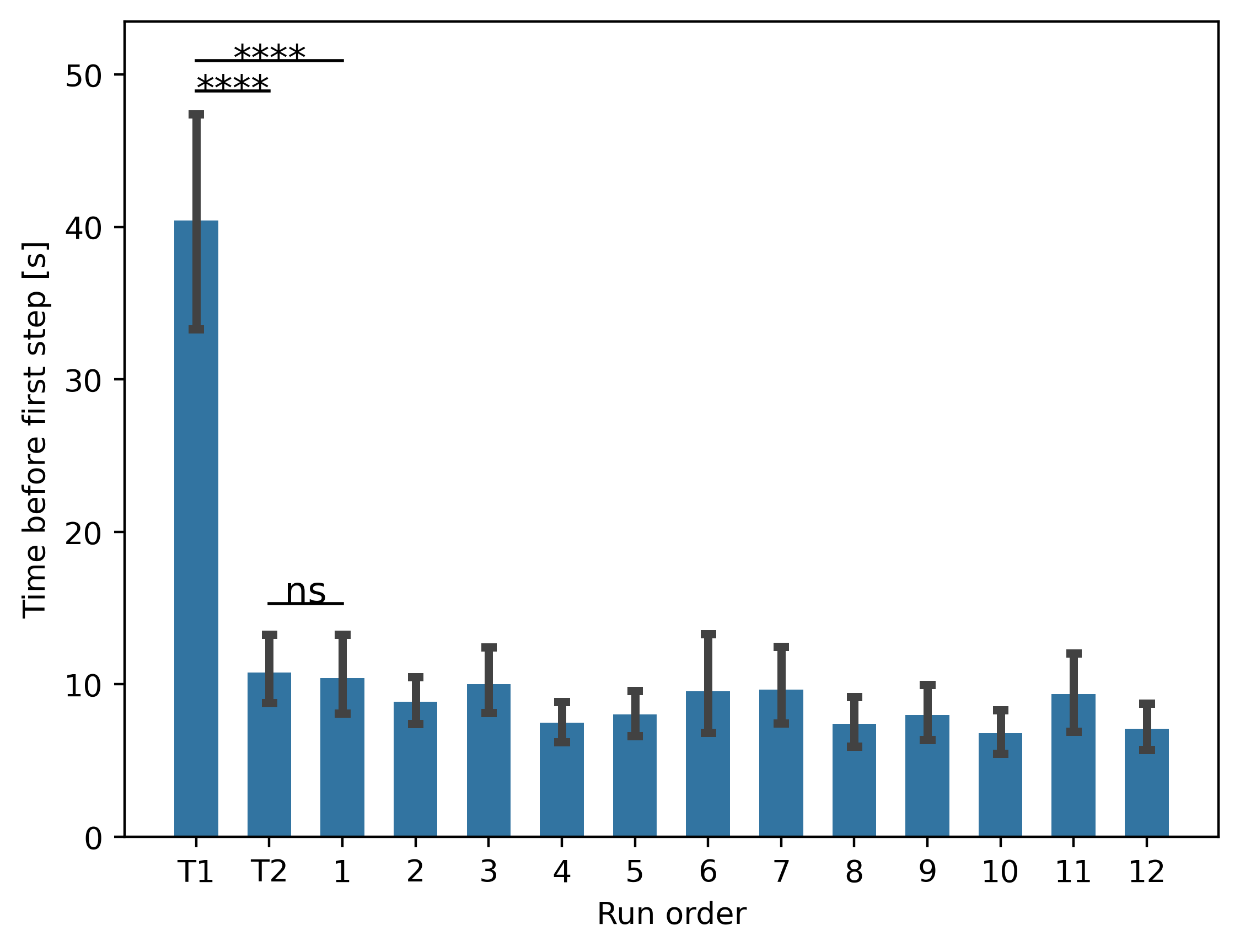}
        \caption{Time before the first step}
        \label{fig:order_tbfs}
    \end{subfigure}\\
    \textit{ns $p > 0.05$, $* p \leq 0.05$, $** p \leq 0.01$, $*** p \leq 0.001$, $**** p \leq 0.0001$}
    \caption{Metrics as function of the run order.}
    \label{fig:order_variables}
\end{figure}

As calculated in Table \ref{tab:Kruskal-Wallis_variable}, the evaluation run order (from 1 to 12) shows no significant difference across the three variables, indicating that performing the test 12 times does not improve participant performance in these aspects, ensuring comparability across runs. To assess the impact of the two training courses, labeled "T1" and "T2," we compared participant performance between these training courses and the evaluation runs, as illustrated in Figure \ref{fig:order_variables}. A learning effect is evident in the training courses: T1 consistently exhibits the "worst" performance, whether compared to T2 or the subsequent evaluation runs. This is expected, as T1 was the first course participants completed. Performance improves significantly in T2, and from that point on, there is no significant difference between the subsequent runs. This suggests that two training courses are sufficient to help participants become familiar with the setup without causing a memory effect.

\begin{figure}[!htbp]
    \centering
    \includegraphics[width=0.8\linewidth, valign=c]{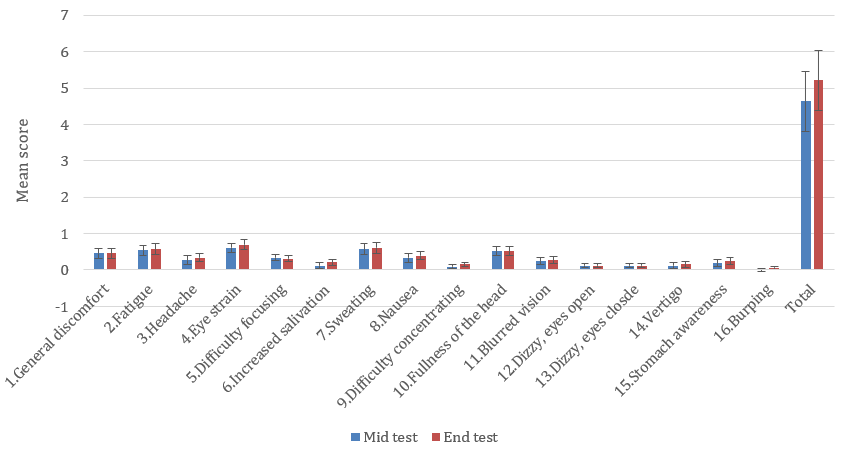}\\
   % \textit{Note: 0=none / 1=slight / 2=moderately / 3=severely} 
    \caption{Average score per item and in total on the SSQ questionnaire for 33 participants, mid-experiment (blue) and end-experiment (red).}
    \label{fig:ssq}
\end{figure}

 Tolerance data, based on responses to the SSQ administered at the midpoint and conclusion of the experiment, are shown in Figure \ref{fig:ssq}. Each item of the SSQ is assessed on a 4-point scale (None, Slight, Moderately, Severely) and the maximum score is 48. All the average scores per item are below 1 and, the total average SSQ score at the midpoint and end of the experiment is between 4.5 and 5.5, which can be considered as negligible or minimal~\cite{Bimberg_Weissker_Kulik_2020}. However, 3 participants reported scores above 11 (12, 15 and 18, respectively) which can be considered as significant and  4 participant prematurely terminated the experiment due to cybersickness (despite SSQ scores below 11). Of these 4 participants, 3 discontinued the experiment after the 8th and the 9th courses. It can be assumed that the intolerance may be due to the duration of the experimental protocol and the repetition of the courses, which are not representative of real use cases in clinics.

%Among the 33 participants who completed the questionnaire, 8 reported a score of 0 at the experiment's midpoint, and 7 reported a score of 0 at the end. This means that 76\% of the participants who completed the questionnaire experienced at least one symptom of cybersickness. No serious side effects were reported. Of the four participants who discontinued the test due to intolerance, two completed the SSQ questionnaire, with scores of 5 and 11. Interestingly, three participants who completed the full assessment had even higher total SSQ scores (12, 15, and 18) than those who discontinued. The maximum possible score on the SSQ is 48.

\subsection{Analysis on missed object}
In this study, we focused on "missed objects " as the primary error metric, and therefore, conducted a detailed analysis based on lighting level of the run; grey level, horizontal and vertical position of the objects; and gaze behavior of the participants. This approach allowed us to investigate potential factors contributing to missed objects and to understand participants' behavior during the test.

%We first extracted information about all missed objects, using the features outlined in Table \ref{tab:feature_err}. Basic details such as the object's name and grey level were directly available in the dataset, while implicit behavioral information was derived from the interaction logs and motion data. This approach allowed us to investigate potential factors contributing to missed objects and to understand participants' behavior during the test.

For this analysis, the following indicators were computed for each missed object:
\begin{itemize}[itemsep=0.1em]
    \item Number of gazes: number of 'gaze in' and 'gaze out' pairs in the events.
    \item Gaze duration [s]: sum of the time differences between each 'gaze in' and 'gaze out' pair. 
    \item In-FOV duration [s]: time duration for which a missed object was within the field of view. 
\end{itemize}
Gaze in and gaze out events are extracted directly from interaction logs based on Tobii XR SDK eye tracking data. Gaze in (or out) means that the gaze vector enters (or exits) the object. 

\begin{figure}[!htbp]
  \subcaptionbox{Number of missed objects per object grey value and lighting level. \label{fig:rgb_lumi_count}}%
  {\includegraphics[width=0.3\linewidth]{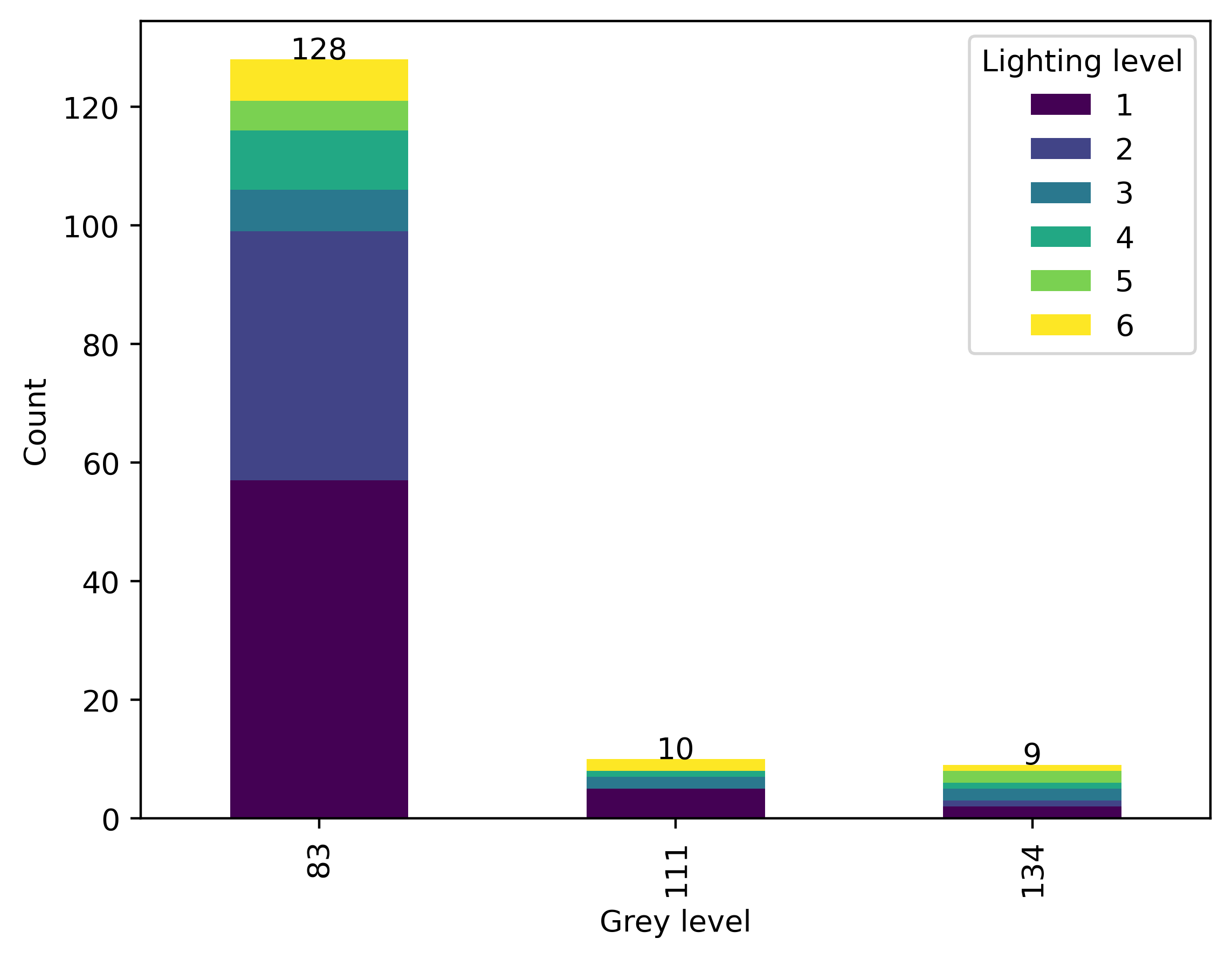}}
  \hspace{\fill}
  \subcaptionbox{Number of missed objects per number of gazes and lighting level.\label{fig:nbgaze_lumi_count}}%
  {\includegraphics[width=0.3\linewidth]{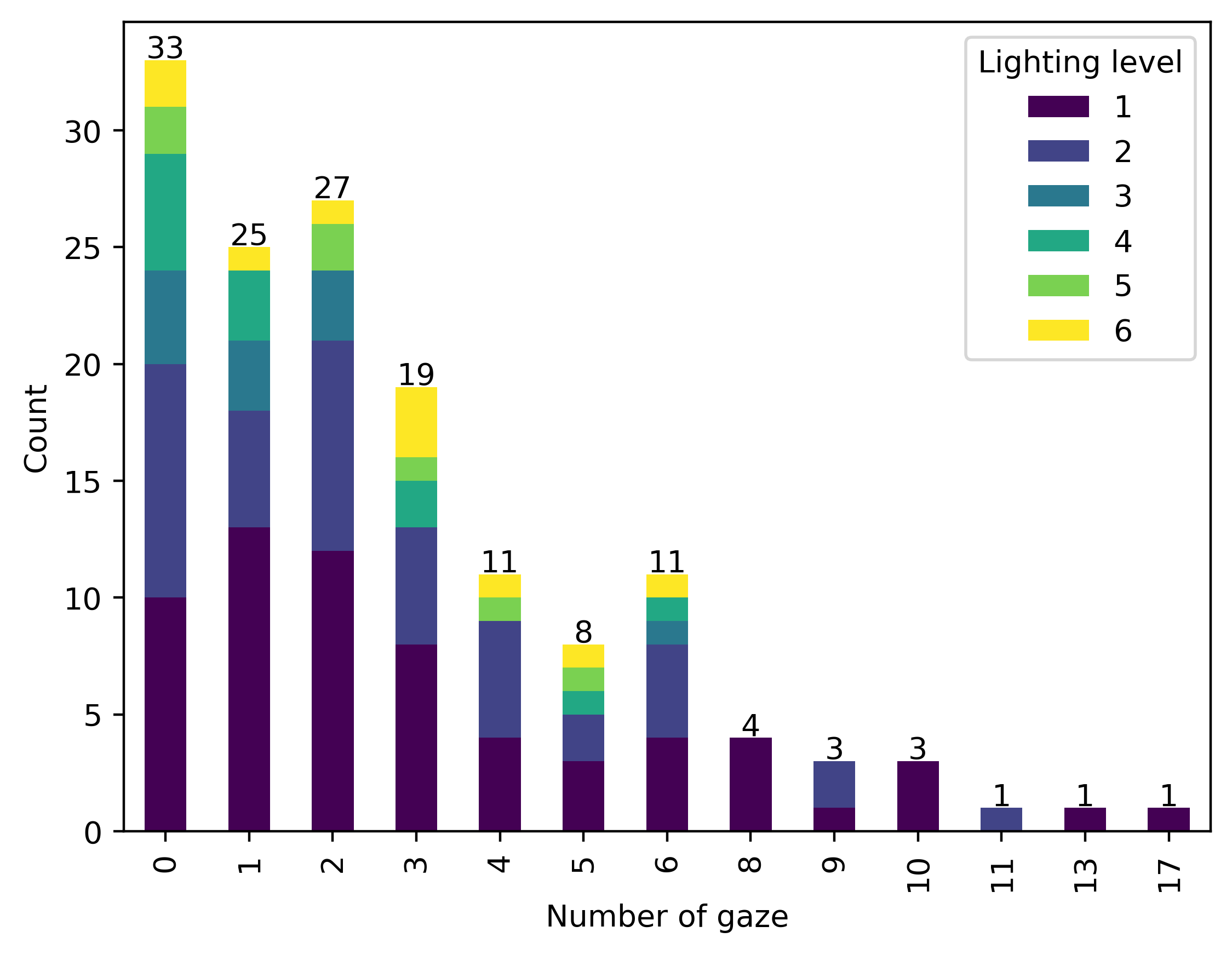}}
  \hspace{\fill}
  \subcaptionbox{Number of missed object per number of gazes and object grey value.\label{fig:nbgaze_rgb_count}}%
  {\includegraphics[width=0.3\linewidth]{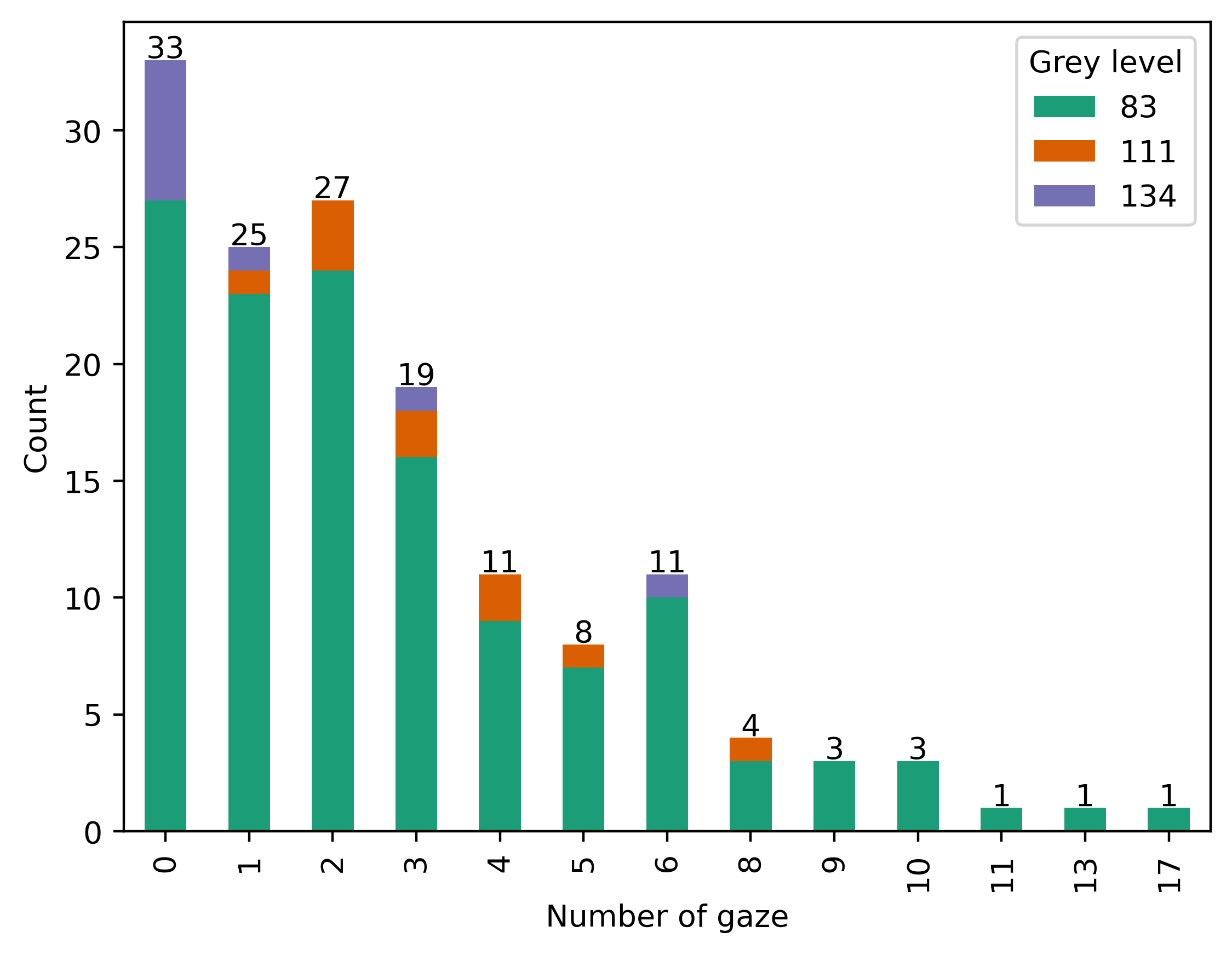}}
  \subcaptionbox{Number of missed objects per object vertical and horizontal position.\label{fig:vert_hori}}%
  {\includegraphics[width=0.3\linewidth]{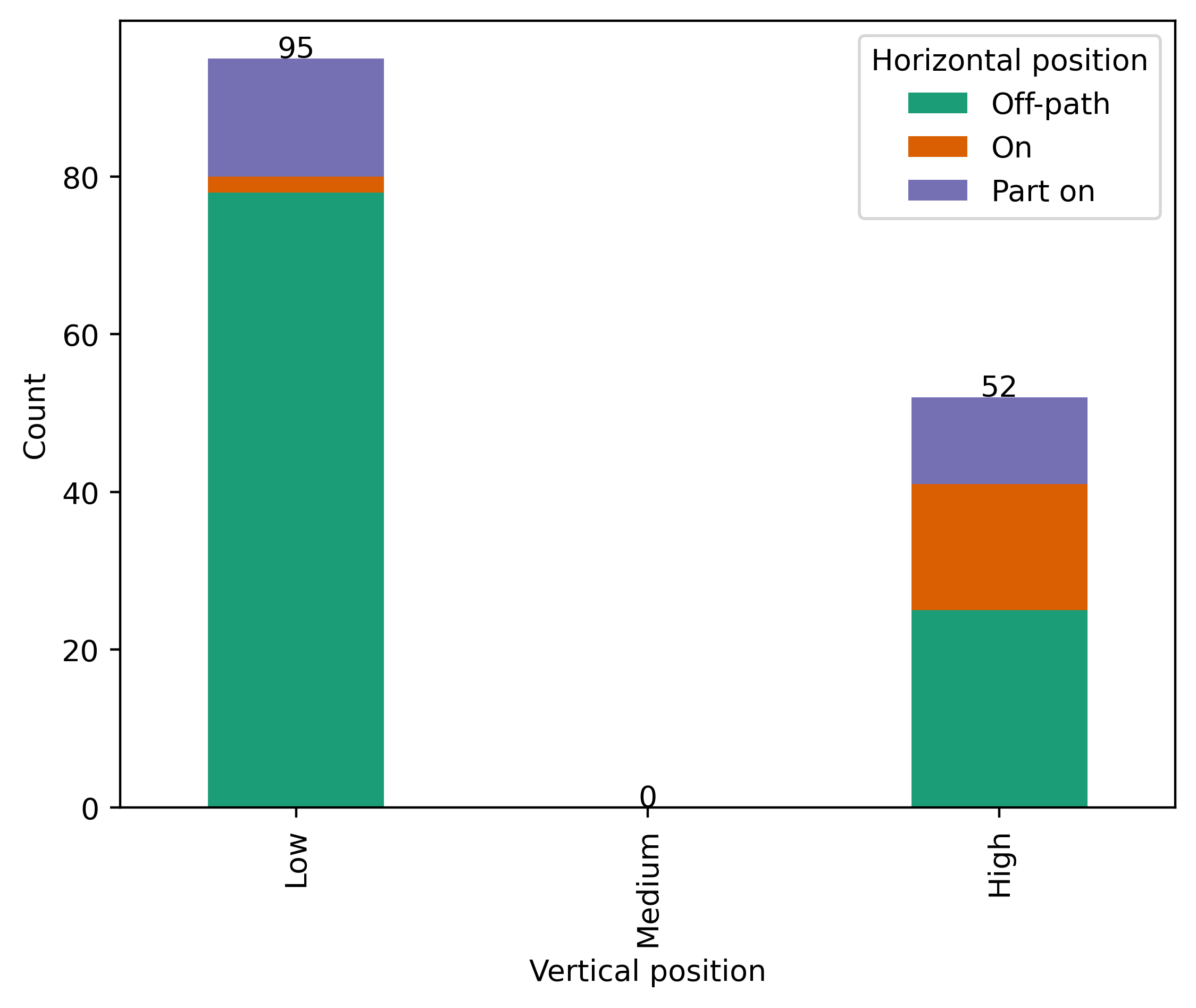}}
  \hspace{\fill}
  \subcaptionbox{Number of missed objects per object grey value and vertical position\label{fig:rgb_vert}}%
  {\includegraphics[width=0.3\linewidth]{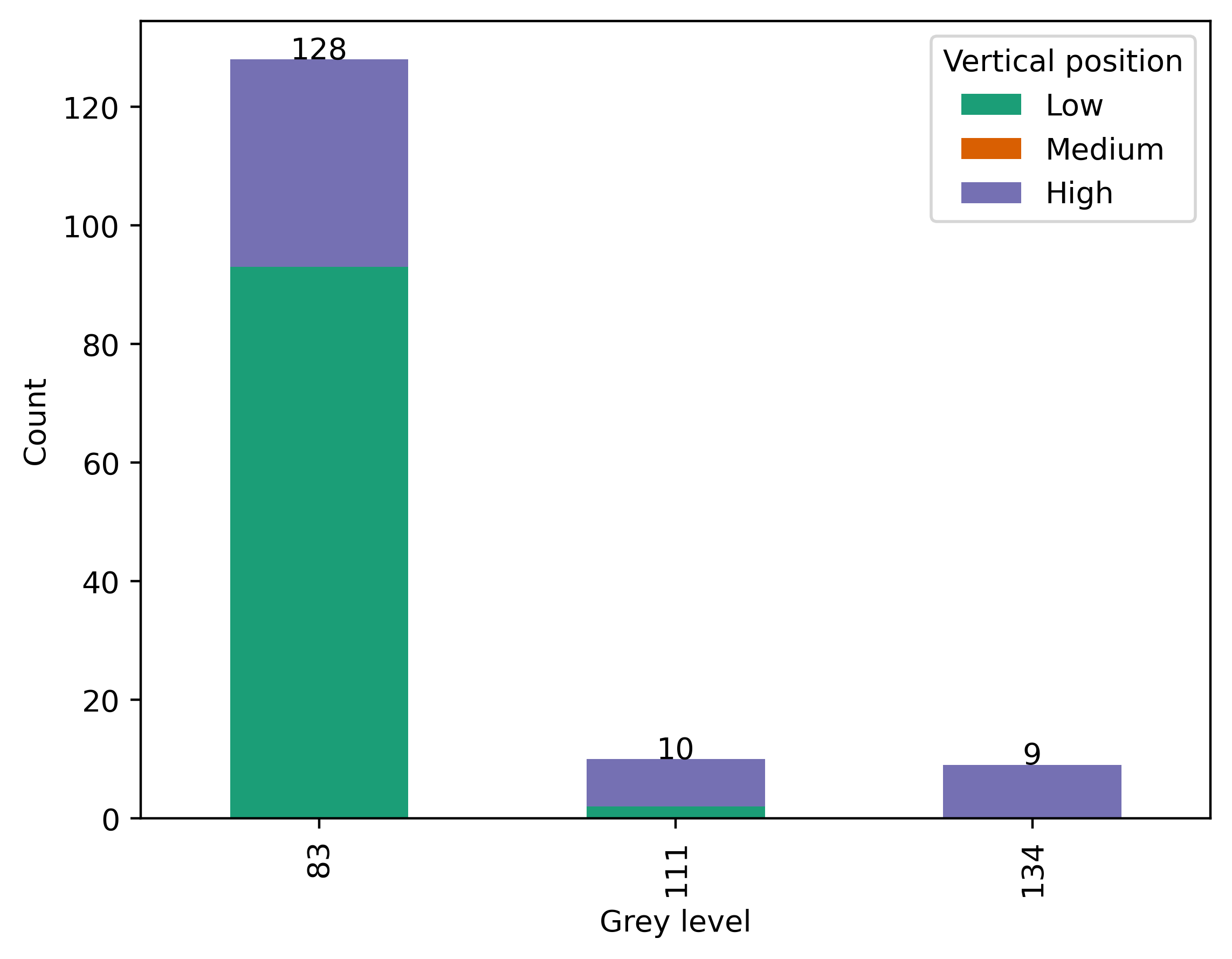}}
  \hspace{\fill}
  \subcaptionbox{Number of missed objects per object grey level and horizontal position\label{fig:rgb_hori}}%
  {\includegraphics[width=0.3\linewidth]{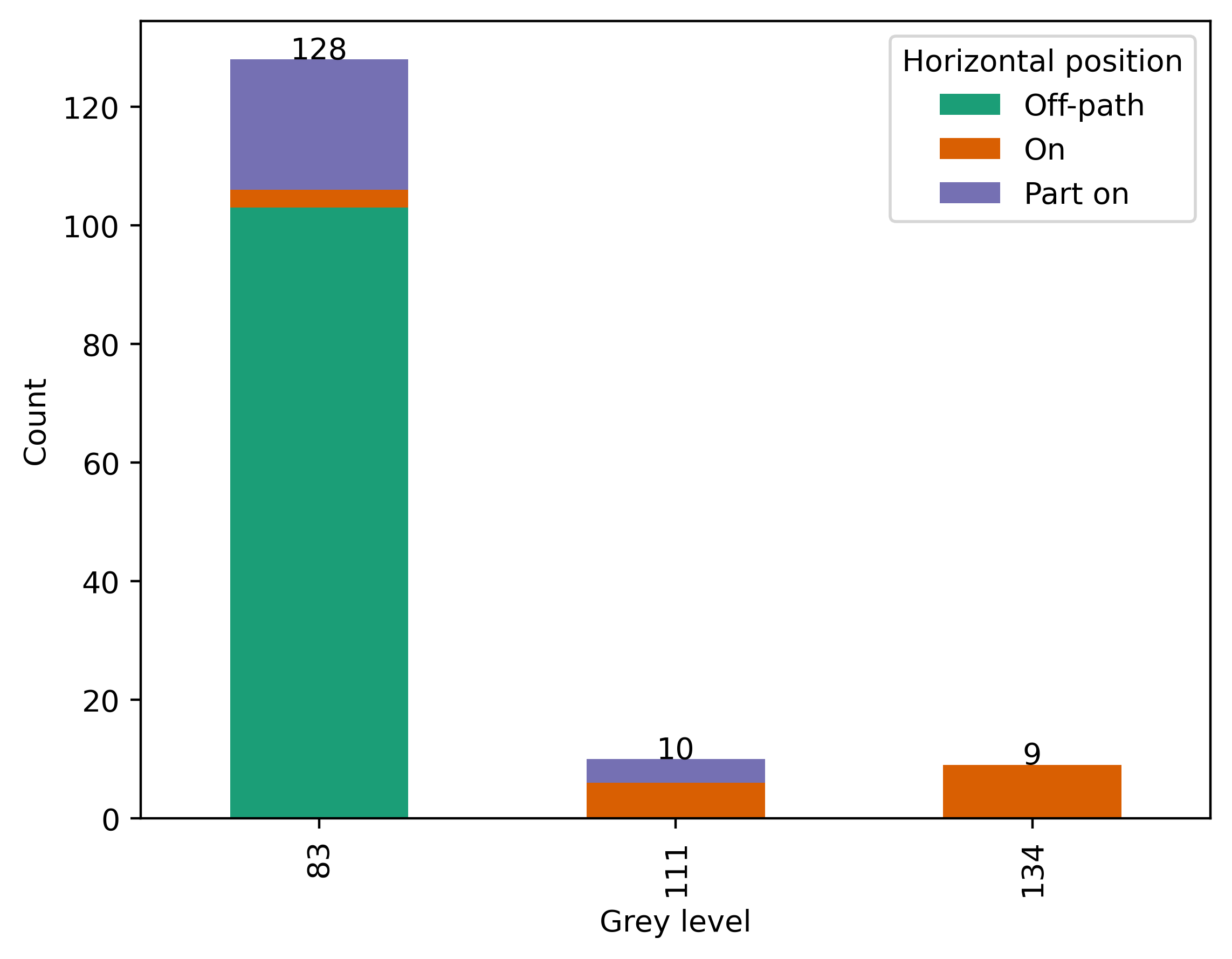}}
  
  \caption{Number of missed objects (Y-axis "Count") for different features.}\label{fig:err_feature}
\end{figure}

As shown in Figure \ref{fig:err_feature}, we analyzed number of missed objects for different characteristics of the environment (lighting level, objects' grey value, objects' position) and the relationship between missed objects and number of gazes. 
%which was a common occurrence in our study (gaze number $\leqslant$ 8, 91.4\%). This trend was consistent across all lighting levels and grey levels. Objects missed despite some gazes (8 $\leqslant$ gaze number $\leqslant$ 13, 7.9\%) often indicate that these objects were difficult to detect but could be critical for evaluating visual performance in more challenging scenarios. These errors primarily occurred in dim environments and with objects of lower grey levels. In contrast, objects missed despite numerous gazes (13 $<$ gaze number, 0.7\%) are considered abnormal errors, possibly due to improper use of the HMD setup. This was rare in our study and occurred only with one particularly dark object in the dimmest environment. It’s important to note that the gaze number thresholds are provided as examples. Future research should delve into more detailed analyses to better understand the underlying causes of these errors.

Figure \ref{fig:err_feature}\subref{fig:rgb_lumi_count} shows that objects with the lowest grey value get easier to be missed, the dim environment caused the miss as well. For objects with the darkest color, they have difficulties to be observed in all lighting levels, brighter environment is helpful for participants to detect dark objects. However, for objects with higher grey levels, they were missed in all lighting levels more equally, which indicates that brighter environment is not helpful for these objects to be detected. These errors can be attributed to participants' careless or improper use of the setup.
Both Figure \ref{fig:err_feature}\subref{fig:nbgaze_lumi_count} and Figure \ref{fig:err_feature}\subref{fig:nbgaze_rgb_count} illustrate that a lower number of gazes correlates with a higher frequency of missed objects. For healthy participants, missed objects with no or very few gazes (1-2) may be due to the object not being scanned at all. These errors are more likely to be due to object position than to grey or lighting levels. We can observe that objects missed despite a significant number of gazes (above 8) are present only in low lighting and grey levels, indicating that these objects were difficult to detect in these conditions and could be critical to observe for low-vision patients in higher lighting levels. Further research should delve into more detailed analyses to better understand the underlying causes of missing an object, especially for a number of gazes between 3 and 7, in order to optimize the use of this test in clinics.

Figure \ref{fig:vert_hori} first reveals that none of the objects with the "medium" vertical position were missed. For objects positioned low on the ground, we find that the farther they are from the path, the more likely they are to be missed. Specifically, most of the missed objects are located off the path, some are partially on the path, and a very small proportion are directly on the path. This suggests that for an object at high position, horizontal placement is not a major factor contributing to be missed. However, for a small object on the ground, its horizontal position influences more how easily participants can detect it. 
Figure \ref{fig:rgb_vert} and Figure \ref{fig:rgb_hori} further illustrate that among the missed objects with the darkest grey level (83), a large proportion were placed at a low position. Similarly, most of the objects were positioned off the path, and as the distance to the path decreased, fewer objects were missed. Indeed, the contrast of dark objects to the path is higher than to the ground.
In contrast, missed objects with higher grey levels (111, 134) were almost exclusively placed at high positions and on the path.
These results can be used to set up a test that optimizes the features (position and color) of the objects, so that only difficult objects are retained in terms of detection and environment' scanning.

%(the only object with a "low" character corresponded to its "part on" horizontal character.). 
%This suggests that for missed objects with higher grey levels -- which should theoretically be easier to detect-- high placement was a key factor contributing to their omission. Additionally, proximity to the path did not appear to facilitate their detection.

%Gaze time duration is strongly correlated with the number of gazes (Pearson correlation coefficient = 0.734, p $\ll$ 0.001). As shown in Figure \ref{fig:hist_two_time}\subref{fig:hist_timegaze}, most missed objects were viewed for only a short period. Figure \ref{fig:hist_two_time}\subref{fig:hist_timeFOV} shows the distribution of the time duration during which missed objects were within the field of view (FOV) over a 2-second window. The majority of missed objects were visible for between 0.5 and 1.5 seconds out of a potential 2.2 seconds. Instances where objects were either completely out of the FOV or constantly visible were rare. We did not identify any significant relationship between this time duration and other features. Further analysis is needed to incorporate the size of each object into the calculations for a more comprehensive understanding.

%%%%%%%%%%%%%%%%%%%%%%%% Continue %%%%%%%%%%%%%%%%%%%%%%%%

%\newpage
\section{Discussion}
%Potential uses for the dataset
%Comparison with other public datasets and harmonization with data dictionaries of similar work
In this study, we introduce a novel virtual reality seated orientation and mobility test protocol along with a dataset collected from healthy subjects within our virtual environment. Preliminary analyses have been conducted based on this dataset to explore the potential applications and insights that can be drawn from the experimental protocol.

% 1. diff compared to MLMT
The design of our virtual environment was largely inspired by the Multi Luminance Mobility Test (MLMT) \cite{chung2018novel}, a physical orientation and mobility test that can be performed under varying lighting conditions. A significant difference between the two is that our tests are conducted entirely within a virtual environment, offering ease of reproduction and access to more detailed data. The arrangement of objects also differs: in our virtual environment, some objects are placed at higher elevations to challenge the upper visual field and encourage exploration of the entire space. In contrast, the MLMT places all objects on the floor at different heights, but no higher than the participant's waist.

%2. diff compared to two vr test
Several VR-based O\&M tests have been developed as well. The VR-O\&M test by Aleman et al. \cite{aleman2021virtual} focused on patients with inherited retinal degenerations (IRDs). This test was validated with 7 healthy subjects and 3 patients, demonstrating the effectiveness of using VR for such assessments. A more recent optimized version of this VR-O\&M test \cite{bennett2023optimization} was evaluated with 20 healthy subjects and 29 patients with IRDs, showing reliable and reproducible differentiation between subjects with IRDs and healthy individuals. These previous VR-based O\&M tests provide a solid proof of concept for the development of our innovative protocol. Among the O\&M tests mentioned (MLMT and the two versions of VR-O\&M), object recognition evaluation mechanisms vary. In the MLMT and the initial VR-O\&M, "obstacle avoidance" was used to determine whether objects were detected. This approach works well in physical tests because participants can feel the sensation of touch. However, in virtual reality, this sensation is absent, although it is possible to add haptic feedback, such as a controller vibration upon object collision. To ensure that participants have truly seen the object in our study, we opted for an "obstacle destruction" mechanism, similar to the "obstacle tag" system used in the optimized version of the VR-O\&M test \cite{bennett2023optimization}. It is worth noting that the VR-S-O\&M test protocol we are presenting was developed prior to the publication of this recent optimization version.

%3. diff & strength: seated displacement
One of the key strengths of the O\&M test we are introducing is the seated position. This approach not only conserves space in practical terms but, more importantly, eliminates variables associated with the physical movement of the patient. Visually impaired patients often have a cautious gait and a fear of falling, which, in a traditional O\&M test involving physical displacement, could result in test failure due to non-visual factors. Additionally, our seated test design makes it possible for patients with motor deficits to participate, provided they can operate a finger trigger and turn a rotating chair. This setup allows for a more inclusive and comprehensive assessment.
% why seated displacement
We opted for continuous virtual displacement to more closely replicate realistic visual stimulation. It is well-known that continuous displacement while seated is more likely to induce cybersickness \cite{ang2023reduction} than physical movement of the subject with an HMD, which would require more space, or teleportation-based displacement, which could detract from the realism we aim to evaluate. The choice of a rotating chair offers a compelling compromise, maintaining the seated position while preserving intuitive, intentional movement for orientation, thereby enhancing immersion. 

% another detail about how to choose obj
Additionally, we designed an environment that strikes a balance between being immersive enough for participants to engage with the virtual reality while remaining non-figurative to avoid visual recognition by inferring familiar environments or objects. In the latest version of the virtual orientation and mobility test by Bennett et al. \cite{bennett2023optimization}, objects resembling everyday items (such as tables, chairs, pendulums, doors, etc.) were used. While this choice enhances immersion, it also raises concerns about visual recognition through deduction from familiar shapes, potentially influencing test outcomes.

% The novelty in providing a dataset
In addition to introducing a new VR-based O\&M test protocol, this study significantly enriches the available datasets dedicated to quality of life (QoL) and functional vision evaluation. To our knowledge, it is the first O\&M dataset to include detailed behavioral information, particularly external eye tracking data, offering a valuable resource for further research in this area.

% the diff in data analysis
Based on the dataset we collected, our data analysis offers a distinct perspective compared to previous O\&M tests. The three O\&M tests we reference share a common scoring system, primarily using the time to complete the course and the number of errors to generate scores (time score and accuracy score). This approach makes sense for the physical MLMT test, given the limitations of data they could collect. However, it is somewhat regrettable that in VR-based tests, rich behavioral data have not been fully utilized. Our dataset opens up more possibilities for future research, as it includes extensive behavioral information, particularly external eye tracking data.
In our preliminary analysis, we did not adhere to the previous scoring system because we used a different type of error definition. Instead, we began by analyzing basic variables. Additionally, earlier studies used errors to assign penalties to scores, implying that they attributed all errors to a decline in functional vision. Our findings suggest that this assumption does not fully apply, as our analysis, conducted solely on healthy subjects, shows that they still made mistakes. This indicates that not all errors are caused by functional vision loss. To better understand the causes of these errors, we proposed analyses focusing on the errors, which has not been explored in previous studies.

%%%%%%%% Data analysis result
% 1. luminance threshold effect
We observed a threshold effect in performance metrics (time duration, number of missed objects, and time before the first step) starting from lighting level L3, which aligns with findings from two previous VR-O\&M tests. Direct comparison of our lighting levels with those used in MLMT or other VR-based tests is not feasible due to differences in virtual environments and HMDs. In contrast to the physical tests, our virtual environment features diffuse ambient lighting without a distinct light source, meaning that illumination is consistent throughout the environment and shadows are not projected. In addition, we provide detailed information about the test's lighting characteristics and the calibration of the HMD display, enabling better reproducibility of the test for research and clinical use.
%Virtual reality HMD are being used more and more in research, and there is a great lack of standardisation in terms of examination conditions and screen calibration. A method for accurately qualifying HMD screens should be develop.

We found that while there is no significant difference in time duration and time before the first step across different courses, there is a notable difference in the number of missed objects (with each course being tested approximately 80 $\pm$ 1.5 times). This result is contrary to our expectations, as the virtual environment was designed to vary the environment while maintaining consistent difficulty. The findings suggest that certain course configurations affect the difficulty of object recognition. Further analysis is needed in future work to identify the factors contributing to this discrepancy.

The aim of the pre-training is to minimize cognitive load by focusing solely on assessing functional vision, rather than the ability to use the virtual reality device. This justifies our protocol of administering two complete training courses before starting the evaluations. Our analysis reveals a significant learning effect, particularly between the first training course (T1) and the second (T2). This indicates that two training courses are sufficient for healthy subjects to become proficient with the setup, allowing them to focus on the functional vision assessment during the actual evaluation.

Analyzing errors made by healthy subjects helps us distinguish between errors unrelated to functional vision, which can be useful for future analyses involving patients. Based on object features and behavioral information, we propose that missed objects can generally be categorized into different groups: errors due to lack of visual analysis of the environment, errors due to difficulty in detection and errors due to misuse of setup. Further analysis is needed to validate these categories and refine our understanding of how different types of errors impact the assessment of functional vision.

%Summary
In summary, this article details the development of a virtual reality-based orientation and mobility test designed to assess functional vision under varying light conditions. The key advantages of our test protocol over previous methods include its reproducibility and configurability due to its VR implementation and lighting calibration, the safety of conducting tests in a seated position, and the comprehensive data it provides on interactions and motion. We introduced a dataset that captures the interactions and behaviors of healthy subjects during our virtual reality seated O\&M test (VR-S-O\&M test). This dataset includes detailed information on subjects, test implementation, and interactions recorded during the experiment. Based on the dataset, we conducted several preliminary analyses. The previous scoring system is not suitable for our test due to its different virtual environment. To establish ecological validity, we focused on basic variables to develop our O\&M performance metrics, such as time duration and the number of missed objects. Our analyses revealed some common phenomena: a threshold effect with varying lighting levels and a learning effect between training and formal runs, which demonstrate the feasibility and validity of our protocol. We also discovered new findings: performance varied across different test courses, indicating that certain configurations might alter difficulty levels. Additionally, our analysis of missed object features showed that errors could result from various factors, such as carelessness, difficulty in observation, and misuse of the setup, rather than solely from declines in functional vision.

% future
This study is therefore an encouraging proof of concept that highlights the need to go beyond traditional statistical metrics and delve into behavioral analysis to create a more nuanced and robust behavior-based scoring system. We recognize that further work is necessary to understand how different configurations affect participant performance, to enhance the quality of virtual graphics for a better immersive experience, and to continue evaluations with visually impaired subjects to compare their behaviors with those of healthy individuals. Virtual reality proves to be a promising tool for the quantitative assessment of functional vision in visually impaired patients, offering numerous possibilities for developing immersive environments and new ecological and interaction variables for longitudinal disease monitoring and clinical trials.

\section*{Dataset availability}
The dataset described in this paper, along with code for data cleaning and preprocessing, is available on Zenodo: \url{https://zenodo.org/records/14918781}. The Unity-based virtual reality project can be provided upon request by contacting \url{toinon.vigier@univ-nantes.fr}.

\section*{Acknowledgments}
This work was supported by the Institut Universitaire de France (IUF) and the Fédération des Aveugles et Amblyopes de France (FAAF). We gratefully acknowledge their support, which made this research possible. We would also like to thank Lucas Communier for his contributions to the development of the system during his internship, which was supported by the NExT initiative (FAME cluster). The NExT initiative receives funding from the French State, managed by the National Research Agency (ANR), under the France 2030 program (reference ANR-16-IDEX-0007), and financial support from Nantes Métropole, the Pays de la Loire Region, and the European Union (FEDER).

\bibliographystyle{ama}
\bibliography{main_tvst}

\newpage
\appendix
\section{Characteristics of objects}
\label{app:objs}
The following table (Table~\ref{tab:objectscharacteristics}) presents the characteristics of the objects all along the path of each course (A to F).  

\begin{table}[H]
\centering
\fontsize{10pt}{10pt}\selectfont
\begin{tabular}{llll}
\hline
Obj Name      & Grey Level & Vertical Position & Horizontal Position \\ \hline
A\_cylinder 1 & 134        & Low               & On the path                 \\
A\_pyramid 2 & 83         & Low               & Partially on the path               \\
A\_cube 1     & 83         & Medium           & On the path                 \\
A\_cube 2     & 83         & High              & Partially on the path               \\
A\_cylinder 2 & 134        & Medium           & Partially on the path               \\
A\_pyramid 0 & 111        & Medium           & Off the path                \\
A\_cylinder 0 & 83         & High              & Off the path                \\
A\_pyramid 1 & 111        & High              & On the path                 \\
A\_cube 0     & 83         & Low               & Off the path                \\ \hline
B\_cylinder 2 & 134        & Medium           & Partially on the path               \\
B\_pyramid 1 & 83         & High              & On the path                 \\
B\_cube 0     & 134        & Low               & Off the path                \\
B\_cube 2     & 83         & High              & Partially on the path               \\
B\_pyramid 2 & 134        & Low               & Partially on the path               \\
B\_cylinder 0 & 83         & High              & Off the path                \\
B\_cylinder 1 & 111        & Low               & On the path                 \\
B\_cube 1     & 111        & Medium           & On the path                 \\
B\_pyramid 0 & 134        & Medium           & Off the path                \\ \hline
C\_cylinder 0 & 83         & High              & Off the path                \\
C\_pyramid 0 & 134        & Medium           & Off the path                \\
C\_pyramid 2 & 134        & Low               & Partially on the path               \\
C\_cylinder 2 & 134        & Medium           & Partially on the path               \\
C\_pyramid 1 & 134        & High              & On the path                 \\
C\_cube 2     & 111        & High              & Partially on the path               \\
C\_cube 1     & 111        & Medium           & On the path                 \\
C\_cube 0     & 83         & Low               & Off the path                \\
C\_cylinder 1 & 111        & Low               & On the path                 \\ \hline
D\_pyramid 1 & 134        & High              & On the path                 \\ 
D\_pyramid 2 & 83         & Low               & Partially on the path               \\
D\_cube 0     & 134        & Low               & Off the path                \\
D\_pyramid 0 & 134        & Medium           & Off the path                \\
D\_cylinder 1 & 134        & Low               & On the path                 \\
D\_cube 2     & 83         & High              & Partially on the path               \\
D\_cylinder 2 & 111        & Medium           & Partially on the path               \\
D\_cube 1     & 111        & Medium           & On the path                 \\
D\_cylinder 0 & 83         & High              & Off the path                \\ \hline
E\_cube 1     & 134        & Medium           & On the path                 \\
E\_cylinder 2 & 134        & Medium           & Partially on the path               \\
E\_pyramid 0 & 134        & Medium           & Off the path                \\
E\_cube 2     & 111        & High              & Partially on the path               \\
E\_pyramid 2 & 83         & Low               & Partially on the path               \\
E\_cube 0     & 83         & Low               & Off the path                \\
E\_cylinder 0 & 83         & High              & Off the path                \\
E\_cylinder 1 & 111        & Low               & On the path                 \\
E\_pyramid 1 & 111        & High              & On the path                 \\ \hline
F\_cube 0     & 83         & Low               & Off the path                \\
F\_cylinder 1 & 111        & Low               & On the path                 \\
F\_cylinder 0 & 83         & High              & Off the path                \\
F\_cube 1     & 134        & Medium           & On the path                 \\
F\_pyramid 1 & 134        & High              & On the path                 \\
F\_pyramid 2 & 83         & Low               & Partially on the path               \\
F\_cube 2     & 111        & High              & Partially on the path               \\
F\_cylinder 2 & 134        & Medium           & Partially on the path               \\
F\_pyramid 0 & 134        & Medium           & Off the path                \\ \hline
\end{tabular}
\caption{Characteristics of the objects for each course (A to F).}
\label{tab:objectscharacteristics}
\end{table}

\newpage
\section{Lighting calibration and measures}
\label{app:light}
The calibration of the HMD Vive Pro Eye was done with the colorimetric probe SpyderX Elite. We measured luminance in cd/m2 per eye for displayed grey value of [0, 31, 63, 95, ..., 255]. The calibration curves are presented below in Figure~\ref{fig:hmd_calibration}. 

\begin{figure}[!htbp]
    \centering
    \includegraphics[width=0.8\textwidth]{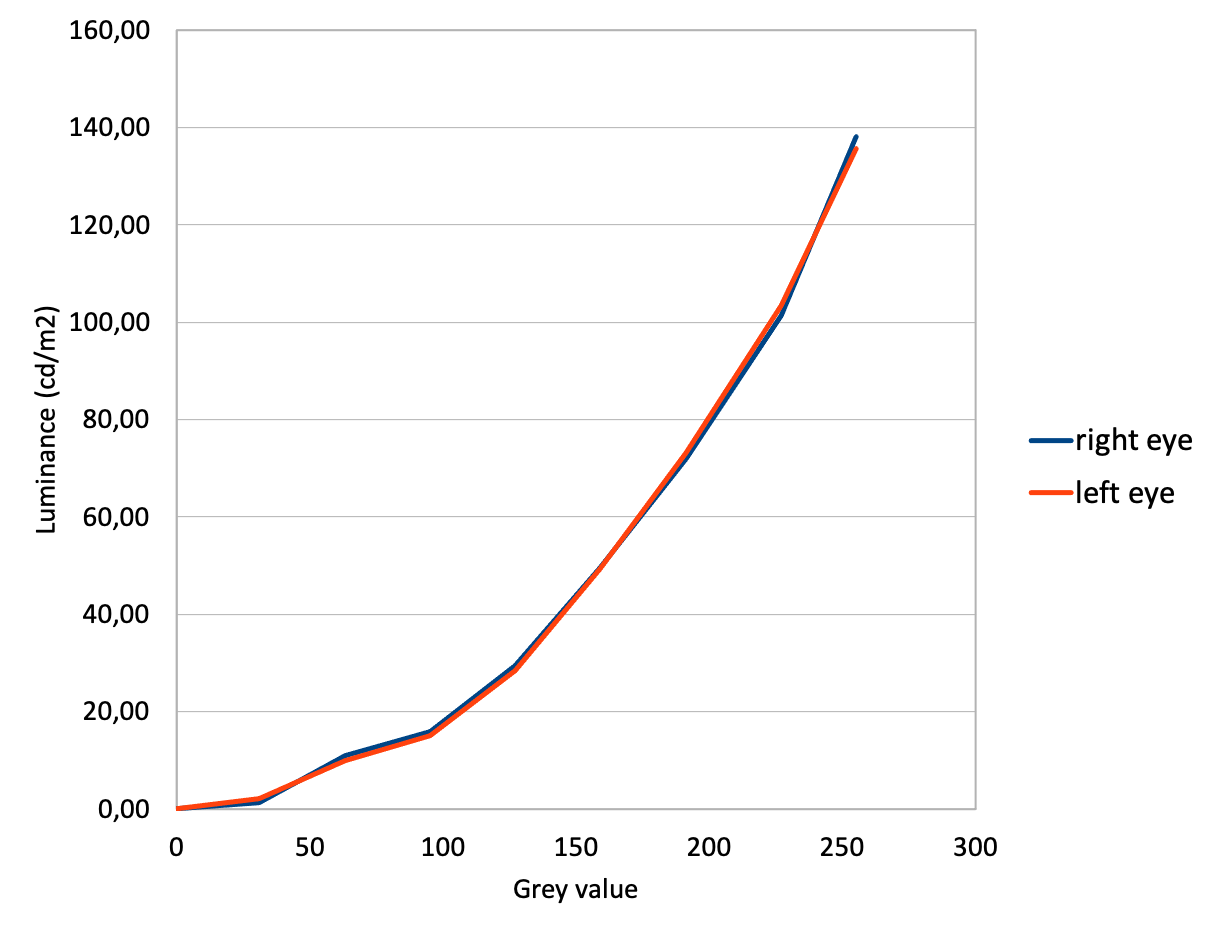}
    \caption{Luminance calibration curves per eye for the HMD Vive Pro Eye.}
    \label{fig:hmd_calibration}
\end{figure}

The intensity value of the ambient light in Unity is set as presented in Table~\ref{tab:ambientlight_values} for each lighting level. 
Using the default graphical pipeline in the motor engine for rendering the virtual scene, we then measured the grey value of each element of the virtual environment for each lighting level (Table~\ref{tab:objects_greyvalue}) and, using linear interpolation from calibration curves, we estimated the displayed luminance of these virtual elements in the HMD used for the experiment presented in this paper (Table~\ref{tab:objects_luminance}).

\begin{table}[!htbp]
\centering
\begin{tabular}{llllll}
\hline
L1 & L2 & L3 & L4 & L5 & L6 \\ \hline
23 & 33	& 55 & 85 & 128 & 175\\
\hline
\end{tabular}
\caption{Ambient light values in Unity for the 6 lighting levels of the test.}
\label{tab:ambientlight_values}
\end{table}

\begin{table}[!htbp]
\centering
\begin{tabular}{lccccccccc}
\hline
 & \textbf{Clear} & \textbf{Medium} & \textbf{Dark} & \textbf{Path}	& \textbf{Floor}	& \textbf{Arrows} & \textbf{Walls} & \textbf{Ceiling} \\ \hline
\textbf{Material} & 134 & 111 & 83 & 101 & 79 & 83 & 155 & 176\\
\textbf{L1} & 6	& 2	& 0	& 2	& 0	& 1	& 5	& 1 \\
\textbf{L2}	& 10 & 7 & 2 & 6 & 1 & 3 & 10 & 5 \\
\textbf{L3}	& 21 & 16 & 9 & 15 & 8 & 11 & 20 & 13 \\
\textbf{L4} & 36 & 28 & 19 & 26 & 17 & 21 & 34 & 24 \\
\textbf{L5} & 58	& 47 & 33 & 44 & 31 & 35 & 56 & 39 \\
\textbf{L6} & 83 & 67 & 49 & 63 & 45 & 52 & 81 & 58 \\
\hline
\end{tabular}
\caption{Grey value of the diffuse material of the virtual elements in Unity and of the same rendered elements for the 6 lighting levels. (Clear, Medium, Dark represent the 3 different grey objects along the path in the test.)}
\label{tab:objects_greyvalue}
\end{table}

\begin{table}[!htbp]
\centering
\begin{tabular}{lccccccccc}
\hline
 & \textbf{Clear} & \textbf{Medium} & \textbf{Dark} & \textbf{Path}	& \textbf{Floor}	& \textbf{Arrows} & \textbf{Walls} & \textbf{Ceiling} \\ \hline
\textbf{L1}	& 0.438	& 0,.216 & 0.105 & 0.216 &	0.105 &	0.160 &	0.382 &	0.160 \\
\textbf{L2}	& 0.660	& 0.493 & 0.216	& 0.438	& 0.160 & 0.271 & 0.660	& 0.382 \\
\textbf{L3}	& 1.27 & 0.998 & 0.604 & 0.937 & 0.549 & 0.715 & 1.21 & 0.826 \\
\textbf{L4}	& 3.17 & 1.66 & 1.16 & 1.55 & 1.05 & 1.27 & 2.63 & 1.44 \\
\textbf{L5}	& 9.11 & 6.14 & 2.36 & 5.33 & 1.82 & 2.90 & 8.57 & 3.98 \\
\textbf{L6}	& 13.7	& 11.1 & 6.68 & 10.5 & 5.60 & 7.49 & 13.3 & 9.67 \\
\hline
\end{tabular}
\caption{Estimated luminance value (in cd/m2) of the rendered virtual elements for the 6 lighting levels. (Clear, Medium, Dark represent the 3 different grey objects along the path in the test.)}
\label{tab:objects_luminance}
\end{table}

\newpage
\section{Data description}
\label{app:data}
\begin{figure}[H]
    \centering
    \includegraphics[width=0.8\textwidth]{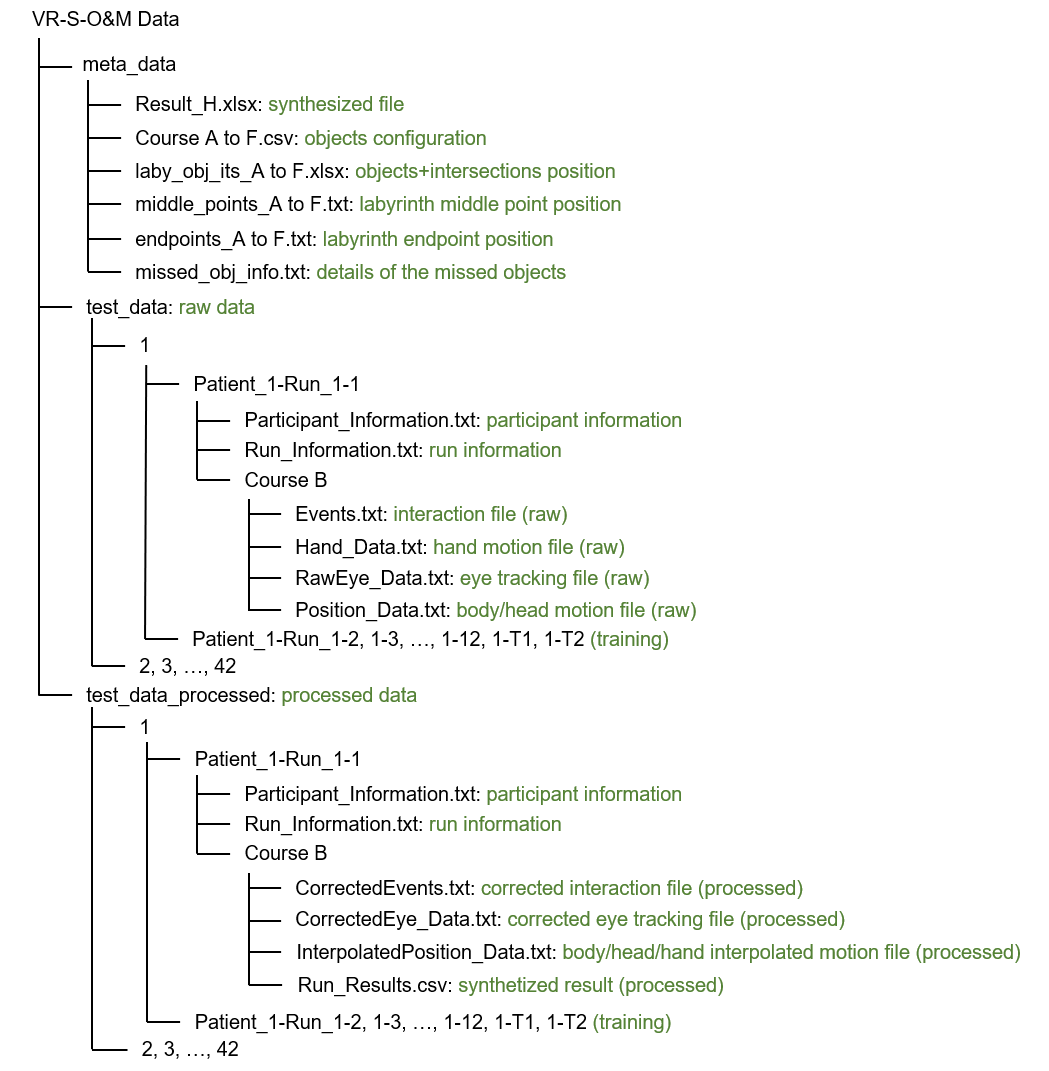}
    \caption{Data structure.}
    \label{fig:data_structure}
\end{figure}

The dataset is housed in a single repository as shown in Figure \ref{fig:data_structure}, which contains three main parts: meta\_data, test\_data and test\_data\_processed.

\subsection{meta\_data:} This folder primarily contains the configuration of the virtual environment along with two files that summarize all runs.

\textbf{Course A to F.csv:} This file contains detailed information about all objects in the environment for each course, including centroid coordinates (x, y, z), scale, gray level, vertical position, horizontal position, and the distance between the bottom of the object and the ground.

\textbf{laby\_obj\_its\_A to F.xlsx:} This file records the coordinates of all objects of interest in each course. Following the direction of the course, it includes information about the start zone, all objects, intersections, and the end zone.

\textbf{middle\_points\_A to F.txt:} This file contains the x and z coordinates of the middle points along the path in each course.

\textbf{endpoints\_A to F.txt:} This file contains the x and z coordinates of the endpoint pairs that define the path.

\textbf{Result\_H.xlsx:} This file provides a summary of the experiment details and results for all runs.

\textbf{missed\_obj\_info.txt} This file summarizes the characteristics of all missed objects across all runs.

\subsection{test\_data} This folder contains the collected raw data without any pre-processing.

\textbf{Participant\_information.txt, Run\_Information.txt:} These files contain basic information about the participant and the run, including participant ID, handedness (left-handed or right-handed), date, and other relevant details. The same goes for those in the test\_data\_processed folder.

\textbf{Events.txt:} This file logs all interactions during each run.

\textbf{Hand\_Data.txt:} This file records hand position time sequences.

\textbf{RawEye\_Data.txt:} This file contains unprocessed eye-tracking data.

\textbf{Position\_Data.txt:} This file stores time sequences of the avatar’s body and head position and rotation.

\subsection{test\_data\_processed} This folder contains the processed data.

\textbf{CorrectedEvents.txt:}  This file contains interaction logs with corrected timestamps. Issues such as missing "system end" entries and redundant logs after "system end" have been resolved.

\textbf{CorrectedEye\_Data.txt:} This file contains eye-tracking data with corrected timestamps, using the system timestamp as a reference.

\textbf{InterpolatedPosition\_Data.txt} This file contains interpolated position and rotation data for the head, hands, and avatar body.

\end{document}